\documentclass[%
 reprint,
 amsmath,amssymb,amsfonts,
 aps,
 prx,
]{revtex4-2}
\usepackage{comment}
\usepackage{amsmath}
\usepackage{amsfonts}
\usepackage{amssymb}

\usepackage{graphicx}% Include figure files
\usepackage{dcolumn}% Align table columns on decimal point
\usepackage{bm}% bold math
\usepackage{xcolor}

\newcommand{\Ulises}[1]{\textcolor{cyan}{UP: #1}}

\begin{document}

\preprint{APS/123-QED}

\title{Forgetting leads to chaos in attractor networks}


\author{Ulises Pereira-Obilinovic}
\affiliation{Department of Statistics, The University of Chicago, Chicago, Illinois, United States of America}
\affiliation{Center for Neural Science, New York University, New York, United States of America}
\author{Johnatan Aljadeff}
\affiliation{Department of Neurobiology, The University of Chicago, Chicago, Illinois, United States of America}
\affiliation{Section of Neurobiology, Division of Biological Sciences, University of California San Diego, La Jolla, California, United States of America}

\author{Nicolas Brunel}
 \email{nicolas.brunel@duke.edu}
\affiliation{Departments of Statistics and Neurobiology, The University of Chicago, Chicago, Illinois, United States of America}
\affiliation{Departments of Neurobiology and Physics, Duke University, Durham, North Carolina, United States of America}

\begin{abstract}

Attractor networks are an influential theory for memory storage in brain systems \cite{hopfield1982neural,amit1985spin,amit1992modeling,brunel2005network}. This theory has recently been challenged by the observation of strong temporal variability in neuronal recordings during memory tasks. 
%Another popular class of recurrent networks for modeling brain dynamics are chaotic networks with random connectivity \cite{sompolinsky1988chaos}. These networks qualitatively recapitulate some features of the heterogeneity observed in recordings in the frontal cortex but lack from associative memory properties.
In this work, we study a sparsely connected attractor network where memories are learned according to a Hebbian synaptic plasticity rule.
After recapitulating known results for the continuous, sparsely connected Hopfield model \cite{tirozzi1991chaos}, we investigate a model in which new memories are learned continuously and old memories are forgotten, using an online synaptic plasticity rule.
We show that for a forgetting time scale that optimizes storage capacity, the qualitative features of the network's memory retrieval dynamics are age-dependent: most recent memories are retrieved as fixed-point attractors while older memories are retrieved as chaotic attractors characterized by strong heterogeneity and temporal fluctuations. Therefore, fixed-point and chaotic attractors co-exist in the network phase space. The network presents a continuum of statistically distinguishable memory states, where  chaotic fluctuations appear abruptly above a critical age and then increase gradually until the memory disappears. We develop a dynamical mean field theory (DMFT) to analyze the age-dependent dynamics and compare the theory with simulations of large networks. We compute the optimal forgetting time-scale for which the number of stored memories is maximized. We found that the maximum age at which memories can be retrieved is given by an instability at which old memories destabilize and the network converges instead to a more recent one. 
Our numerical simulations show that a high-degree of sparsity is necessary for the DMFT to accurately predict the network capacity. %as predicted for networks of binary neurons \cite{derrida1987exactly}.
Finally, our theory provides specific predictions for delay response tasks with aging memoranda. In particular, it predicts a higher degree of temporal fluctuations in retrieval states associated with older memories, and it also predicts fluctuations should be faster in older memories. Our theory of attractor networks that continuously learn new information at the price of forgetting old memories can account for the observed diversity of retrieval states in the cortex, and in particular the strong temporal fluctuations of cortical activity. %We suggest that the statistical mechanics of Hopfield-type networks with online learning and forgetting can account for diversity of retrieval states in the cortex and for the strong fluctuation of cortical activity.

\end{abstract}
\maketitle

\section{Introduction}

Attractor networks are an influential theory based on recurrent networks for learning and memory in brain systems \cite{hopfield1982neural,amit1985spin,amit1992modeling,brunel2005network}. In this theory, memories correspond to stable patterns of network activity representing the stored memoranda, which correspond to fixed-point attractor states of the network dynamics.
% When a memory is learned, changes in the connectivity through synaptic plasticity  driven by an external input to the network produce a distributed connectivity pattern of synaptic modifications. These changes in the connectivity create a fixed-point attractor in the network's dynamics corresponding to the neural representation of the learned memorandum. 
When each memory is learned, the external input associated with the memory modifies the network synaptic connectivity thanks to a synaptic plasticity rule. These synaptic modifications lead to the creation of a fixed-point attractor of the  dynamics, corresponding to a neural representation of the learned memorandum. 
In the attractor state, the network activity is correlated with, but not necessarily identical to, the original external input to the network. 
%The neural representation of the memorandum is latent when the memory is not being retrieved existing as a memory engram in the network's connectivity.  
% Upon an external cue correlated with the stored memorandum that is being retrieved, the network autonomously relaxes to the corresponding attractor state, and the identity of the memorandum can be easily decoded by downstream circuitry. 
Upon a brief presentation of an external cue correlated with one of the stored memoranda, the network retrieves the corresponding memorandum by autonomously relaxing to the associated attractor state, which permits decoding the memorandum's identity by downstream circuitry. The theory parsimoniously reproduces key observations in neuronal responses during memory tasks - in particcular in reproduces the phenomenon of selective persistent activity, i.e.,~ the observation that some neurons in multiple cortical areas have persistent elevated firing rates following the presentation of specific items to be maintained in memory during the task, but not others \cite{fuster1971neuron,miyashita1988neuronal,funahashi1989mnemonic,goldman1995cellular,liu2014medial,guo2014flow,inagaki2017discrete}.

%Typically, this class of models is studied in a scenario where memories are static, in contrast to the  online nature of memory formation and of forgetting in the real world.  Recently, this class of models have been challenged by the strong temporal irregularity and  heterogeneity in time-averaged responses during working memory tasks, which is hard to reproduce in attractor networks with fixed-point memory states.

A major challenge to the attractor network scenario as the mechanism  subserving  mnemonic representations is the observation of high degree of temporal irregularity in neuronal activity during delay periods in the prefrontal cortex \cite{compte2003temporally,shafi2007variability,barak2010neuronal,barak2014working,kobak2016demixed,murray2017stable}. The observed high spike irregularity and non-stationary of the trial-averaged firing rates challenge the classic picture of retrieval states as discrete fixed-point attractor states. Extensions of classic attractor  network models \cite{barbieri2007irregular,mongillo2008synaptic} successfully account for the observed spike irregularity. However, these models fail to reproduce the non-stationarity observed in the firing rates. Various models have been proposed to account this non-stationarity by encoding memories using time-varying variables \cite{mongillo2008synaptic,lundqvist2010bistable}, but they cannot reproduce the stable population coding observed in neuronal recordings during memory tasks \cite{murray2017stable} (but see \cite{druckmann2012neuronal}).
%Various models have been proposed to account for this extra variability within the framework of attractor networks \cite{barbieri2007irregular,mongillo2008synaptic,lundqvist2010bistable,mongillo2012bistability,druckmann2012neuronal}. In general, these models face two types of limitations in reproducing features of neuronal recordings. On one hand, for models based on fixed-point attractor states with injected external variability, they typically fail to reproduce the temporal variability and heterogeneity observed in data, unless an unplausible degree of and parameter fine-tuning is made. On the other hand,  models based in transient activity cannot reproduce the stable population coding observed in neuronal recordings \cite{murray2017stable}.  
%{\color{blue}{[YA-- can we say in 1-2 sentences here, without being too technical, what are the main limitations of the cited literature? Bimodal/unrealistic distribution of rates? Large reduction in capacity? Right now it is not clear why the additional model (which is extended and investigated here) is necessary ].}}

Recently, we have proposed an alternative scenario to account for the observed variability in an attractor network whose learning rules are inferred from {\it in vivo} data \cite{pereira2018attractor}. In this scenario,  memories are stored by chaotic attractors (in contrast to fixed-point attractors in classical attractor networks \cite{hopfield1982neural,amit1985spin,tsodyks1988enhanced}). Neural activity is correlated with the stored pattern at all times, yet the dynamics are characterized by strong internally generated temporal fluctuations.  The network displays associative memory properties in spite of strong chaotic fluctuations in neuronal activity. Using a dynamical mean field theory (DMFT)  \cite{sompolinsky1988chaos,crisanti2018path} Tirozzi and Tsodyks predicted the existence of this chaotic associative memory phase in
the sparse version of the Hopfield model \cite{tirozzi1991chaos}.  In both the Tirozzi and Tsodyks model, as well as in networks constrained by {\it in vivo} data \cite{pereira2018attractor}, all 
fixed-point attractor memory states transition to chaos when either a parameter characterizing the overall strength of synaptic connectivity, or the number of stored patterns, reach a critical value. %\Yonatan{At any given time the number of chaotic attractors in our model is $\propto N$. The number of memories at any given age is not $\propto N$ so it may be better to not get into the question of extensive / not-extensive. Maybe we can say the transition is ``global'' instead of ``extensive''?}
%\Ulises{I don't understand what do you mean with this comment.}\Yonatan{Sorry, this comment was misplaced, I now realize I originally put it before the notion of memory age is introduced. What I meant is that once an age of a memory is introduced, the transition to chaos is no longer extensive, because each memory transitions to chaos ``alone'' in its own time. Nevertheless there are O(N) chaotic states. So we'll have to be careful how to describe the transition once age is introduced.}
%Furthermore, consistent of what  Tirozzi and Tsodyks  predicted for the sparse Hopfield model \cite{tirozzi1991chaos}, it has also been found in this model that the capacity  (i.e. the maximum number of memory states the network can store) for chaotic attractors is larger than what is predicted by a static mean field theory (SMFT) for fixed-point attractors as memory states \cite{pereira2018attractor}. However, a theory for chaotic memory states in networks constrained by data is lacking.

In such models, as well as in classical attractor networks \cite{amit1985spin}, when the number of stored patterns is larger than the network capacity (i.e., maximum number of patterns the network can store) they undergo  catastrophic forgetting, and all memories are forgotten at once (i.e., no memory can be retrieved, no matter how close the initial network state is to the stored memory).   Catastrophic forgetting  occurs because of the statistical symmetry between stored patterns. In a large network, when all patterns are identical and independently distributed, as in classical attractor network models \cite{hopfield1982neural,amit1985spin,tsodyks1988enhanced} forgetting one pattern is statistically equivalent to forgetting all.   The recipe for fixing catastrophic forgetting is well known - it consists in introducing an online learning process, in which new memories are written on top of older memories that are gradually forgotten (sometimes called palimpsest models \cite{nadal1986networks}). In this models, each memory has a different age,  which breaks the statistical symmetry between patterns, allowing successful retrieval of the most recent patterns, at the price of a forgetting of  older ones \cite{parisi1986memory, mezard1986solvable, tsodyks1990associative, amit1994learning}. Online learning in attractor networks has been actively studied using networks of binary neurons \cite{parisi1986memory,amit1994learning,fusi2005cascade,fusi2007limits,romani2008optimizing,dubreuil2014memory,huang2011capacity,lahiri2013memory,benna2016computational}, in which memory states are fixed-point attractors. However, the effect of online  learning rules on the dynamics of memory states remains an open question in networks of continuous rate units.  Building such theory is key for contrasting theoretical predictions with firing rates recorded in neurophysiological experiments. Additionally, since networks of rate units can be obtained as approximations of more realistic networks of spiking neurons \cite{ostojic2011spiking}, understating the dynamics and learning in such networks is a critical intermediate step that combines analytically tractability and biological realism.
%\Yonatan{Why is this important? We need to say here that the binary models are less useful if we want to understand temporal fluctuations AND age-dependent memory.}

Here we study an attractor network of continuous rate units, where each memory is initially acquired and then progressively degrades according to an online Hebbian learning rule. We provide a DMFT describing the network dynamics, in the large $N$, sparse connectivity limit, and compare our theory with numerical simulations. Remarkably, we show that in this network  both fixed-point and chaotic attractors memory states co-exist in a large range of forgetting time scales. The pattern's age determines the nature of  its retrieval state: fixed-point for newer patterns, and chaotic attractor for older patterns, leading to a continuum of statistically distinguishable memory states. Analysis of solutions to the DMFT equations reveals the storage capacity (i.e., maximal age at which memories can be retrieved) and the forgetting time-scale that maximizes capacity. We show that storage capacity is given by an instability line, on which old memories become unstable and the network retrieves a newer one instead. 
%However, for old enough memories the network makes mistakes, forgetting the retrieved older memory and remembering a most recent one,  which is not predicted by our DMFT. By studying the stability of our DMFT we correctly predict this erroneous retrieval. As expected, in the limit of no forgetting our theory coincides with the results obtained by Tirozzi and Tsodyks \cite{tirozzi1991chaos}. 
We complement analytical results by comparison with simulations of large networks (up to 10$^7$ neurons). We find that a good agreement between simulations and theory is obtained only for extremely sparse networks, i.e.,  when the average number of connections per neuron scales logarithmically with the network size. However, the transition to chaos is well predicted by our theory for denser networks, i.e.,  when the average number of connections scales as a power law with the network size.
 
%\Yonatan{The abstract and introduction still have significant overlaps and are both quite long. If the introduction remains this long I think we should shorten the abstract.} 

\section{Model}

We consider a recurrent neuronal network composed of $N$ neurons whose input currents are described by analog
variables $h_i$, where $i=1,2,\dots,N$. 
The instantaneous firing rates of neurons are given by the
the input-output single neuron transfer function (or f-I curve) $\phi(x)=\tanh(x)$. This is a unrealistic choice from a neurobiological point of view, but has the advantage of simplicity. The major features presented in this theory are qualitatively preserved in more realistic networks \cite{pereira2018attractor} (see Discussion). 
The synaptic input currents obey the standard current-based version of the 
rate equations
 \begin{equation}
\label{results:model:1}
% \dot{h}_i=-h_i+\sum_{i\neq j}^{N}c_{ij}J_{ij} \phi(h_j)+I_i, 
 \frac{{\rm d}{h}_i(t)}{{\rm d}t}=-h_i(t)+\sum_{i\neq j}^{N}c_{ij}J_{ij} \phi[h_j(t)]+I_i, 
\end{equation}
where $c_{ij}$ represents the `structural connectivity' of the network, which is generated as a sparse random Erd\H{o}s-R\'{e}nyi graph, i.e.,~all $c_{ij}$ are i.i.d.~Bernoulli random variables, $c_{ij}=1$ with probability $K/N$ where $K$ is the average in- and out-degree; $J_{ij}$ is the strength of
the synapse connecting the pre-synaptic neuron $j$ to the post-synaptic neuron $i$; and $I_i$ is an external input to neuron $i$. We assume that $1\ll K\ll N$, which means that the average number of connections is large but much smaller that the network size.  This is a relevant limit for microcircuits in the brain, where each neuron receives a large number of inputs ($K\sim 1,000$s), and connection probabilities are small, of order $\sim$10\% in cortex\cite{mason1991synaptic,markram1997physiology,holmgren2003pyramidal,thomson2007functional,lefort2009excitatory} and 1\% in hippocampus  \cite{guzman2016synaptic}. 
%Note however that DMFT requires that $K$ grows close to the logarithm of $N$ for fully matching the network memory storage capacity, a scaling leading to sparser connectivity than the one observed in local neuronal microcircuits. A question we address below is how well DMFT approximates the dynamics of less sparse networks. 
%\Yonatan{Since we are emphasizing the fact that the logarithmic scaling of sparsity is important, it may be strange to say that this is ``the relevant limit'' cortical networks with 10\% connectivity. Given that cortical neurons have $\sim 10^3$ inputs, the logarithmic scaling clearly doesn't apply. I think that at least for cortical networks we should more modestly say that it's an approximation.} \Ulises{I agree with this above, but I think it is still OK to say that the $1\ll K \ll N$ is a relevant limit since we are not choosing any scaling yet. In fact, Fig 3 shows that for some power law scaling the approximation is still fair. Also this is for fully matching the storage capacity. I added some sentences regarding this.} 

The synaptic connectivity matrix is assumed to have been structured through past presentations of external stimuli (patterns) shown to the network. The stimulus presentation time scale is assumed much longer than the time scale of network dynamics,
and the synaptic connectivity is assumed fixed at the scale of retrieval of specific patterns. We use the variable $u$ to refer to the stimulus presentation times, to avoid confusion with $t$ in Eq.~(\ref{results:model:1}), where $u\in\mathbb{Z}$. External inputs to the network consist in a stream of time-ordered binary {\it patterns} $\{\vec{\eta}^u\}, u\in\mathbb{Z}$. We assume that $\eta_{i}^u$ are i.i.d.~binary random variables, such that $\eta_i^u=\pm 1$ with equal probability. We note that while the stored memories are binary, the network dynamics and therefore the retrieval states are continuous. The patterns $\eta$ can be interpreted as the population firing rates when the dynamics in Eq.~(\ref{results:model:1}) is driven by a strong random external stimulus $I_i$ symmetrically distributed around zero. The synaptic weights are modified by the patterns in {\it one shot} using an online version of the covariance rule \cite{sejnowski1977storing}. According to such a rule, 
% The entries of the vectors $\eta_{i}^s$ represent the firing rate of neuron $i$ at time $s$.
\begin{equation}
\label{results:model:2}
    J^{u+1}_{ij} =\rho J^{u}_{ij}+  \frac{A}{K}\eta^{u}_i\eta^{u}_j.
\end{equation}

 In Eq.~(\ref{results:model:2}), the first term in the r.h.s.~represents a decay term with a forgetting rate $0<\rho<1$ that prevents synaptic weights from growing unbounded, while the second term represents a `Hebbian' synaptic modification that is proportional to the covariance of pre and post-synaptic activity, with a strength parameter $A$. On one extreme, for $\rho=0$ the learned connectivity from previous patterns is forgotten in one time step, while in the case $\rho=1$ there is no forgetting. The case $\rho=1$ is problematic since it leads to an unbounded growth of the weights when the number of presented patterns becomes very large. However, for completeness we will describe the $\rho=1$ scenario, in which a finite number of patterns is presented, in Section \ref{sec:tirozzitsodyks}.

With such a learning rule, the connectivity matrix is given at any given time by
\begin{equation}
\label{results:model:2b}
J^u_{ij}=\frac{A}{K} \sum_{u'=-\infty}^{u-1} \rho^{u-1-u'}\eta^{u'}_i \eta^{u'}_j. 
\end{equation}

Without loss of generality, we focus on a particular time $u$, drop the index $u$ for the connectivity matrix $J_{ij}=J_{ij}^u$, and rewrite $J_{ij}$ as
\begin{equation}
\label{results:model:3}
J_{ij}=\frac{A}{K} \sum_{\mu=0}^{\infty} e^{-\frac{\mu}{\tau K}}\eta^{\mu}_i \eta^{\mu}_j. 
\end{equation}
where $\mu$ is the age of pattern $\eta^{\mu}_i$ (i.e.,~how far in the past pattern $\mu$ was presented), and $\tau$ is a forgetting time constant defined as $\tau=-1/(K\log\rho)$. Note that in the sum in the r.h.s.~of Eq.~(\ref{results:model:3}), we have changed the pattern indices for convenience, such that a pattern of age $\mu$ corresponds to the pattern shown at time $u'=u-1-\mu$.

%\end{widetext}
\section{Dynamic mean field theory\label{sec:DMFT}}

\subsection{Formalism}

We start by deriving the  DMFT for the network model defined by Eqs.~(\ref{results:model:1}, \ref{results:model:3}), in the limit  of infinitely many neurons $N\to \infty$, synapses per neuron $K\to \infty$,
and sparse connectivity $K/N\to 0$  \cite{kree1987continuous,tirozzi1991chaos}. Although the network in Eqs.~(\ref{results:model:1},\ref{results:model:3}) stores binary patterns, i.e., $\eta_i^s \in \{-1,1\}$, here we present a general DMFT for arbitrary $P(\eta)$, with the constraint that $\langle \eta \rangle=0$. This ensures that the average change in connection strength due to
learning of a single pattern is zero, and therefore the synaptic weights average does not grow with the number of stored patterns.  This could be enforced by a homeostatic mechanism that controls the mean changes in the incoming
inputs due to learning \cite{toyoizumi2014modeling,vogels2011inhibitory}. We use similar methods as in refs.~\cite{sompolinsky1988chaos,tirozzi1991chaos}, and anticipate that for some parameters the dynamics of the network will be chaotic. In the above limit, the instantaneous synaptic inputs to each neuron can then be rewritten of its mean $\nu_i$ and random temporal variations around the mean $\zeta(t)$ due to chaotic dynamics, 
\begin{equation}
    \label{results:results:1}
    \sum_{i\neq j}^{N}c_{ij} J_{ij} \phi(h_j(t)) = \nu_i+\zeta_i(t).
\end{equation}

Therefore, the network dynamics in Eq.~(\ref{results:model:1}) becomes statistically equivalent to a stochastic differential equation (SDE) given by
\begin{equation}
\label{results:mft:2}
%\dot{h}_i(t) = -h_i(t) + \mu_i + \zeta_i (t).
\frac{{\rm d}{h}_i(t)}{{\rm d}t} = -h_i(t) + \nu_i + \zeta_i (t).
\end{equation}
where the $\zeta_i$s are uncorrelated stochastic processes, whose autocorrelation will be computed self-consistently in the following.
%In this limit  Eq.~(\ref{results:model:1}) is reduced to
%\begin{equation}
%\label{mft:1}
%\dot{h}_i(t) = -h_i(t) + \mu_i + \zeta_i (t),
%\end{equation}
%where

%\begin{equation}
%\label{mft:2}
%\mu_i = A  \sum_{\mu = 1}^s %\eta_i^{\mu}e^{-\frac{\mu}{\tau K}}m_{\mu}, 
%\end{equation}

%corresponds to the average input current to neurons $i$.  
As in classical mean field theories
for attractor neuronal network models \cite{amit1985spin,tsodyks1988enhanced}, we define the overlaps between network state and the memories stored in the network as
\begin{equation}
    \label{results:mft:4}
    m_{\mu} = \frac{1}{N}\sum_{i=1}^N\eta_i^{\mu}\phi(h_i) = \langle \eta^{\mu}\phi(h)\rangle_{h,\eta},
\end{equation}
%\begin{equation}
%\label{mft:3}
% m_{\mu}\equiv \langle \eta^{\mu} \phi(h)  \rangle_{h,\eta^{\mu}}.
%\end{equation}
where $\langle \dots \rangle_{h,\eta}$ represents an average over the statistics of the stored patterns $\eta$ and the input currents $h$. The network is able to retrieve a pattern stored in memory if it settles in a state in which there is an overlap of order one with the corresponding pattern. This means in particular that the stimulus identity can be retrieved from the network activity by a linear decoder. On the other  hand, if the overlap is zero in the large $N$ limit, then the memory is forgotten and cannot be retrieved. 
%Here the average is over the distribution of the synaptic input current $h$ and of the stored pattern $\eta^{\mu}$. 
We assume that the overlaps do not depend on time  (i.e., $m_{\mu}(t)= m_{\mu}$), which is trivially true for fixed-point attractors, and becomes a good approximation at large $N$ for chaotic attractor memory states. 
%The definition of the overlaps in Eq.~(\ref{results:mft:4}) is a  natural extensionfor analog neurons and nonlinear learning rules of the overlaps used in classic attractor neural network models.
%We assume that the number of retrieved patterns $|R|$ are of order $|R|\sim \mathcal{O}(1)$, and therefore just a finite number of patterns have an non-negligible overlap with neural activity. 
Except when stated otherwise (see Section \ref{sec:amendedDMFT}), we will assume that in a given retrieval state only a single memory is retrieved. With this assumption, the mean synaptic inputs $\nu_i$ to neuron $i$ become 
\begin{equation}
    \label{results:results:5a}
  \nu_i = A   \eta_i^{\mu}e^{-\frac{\mu}{\tau K}}m_{\mu}.
\end{equation}
%where 
%In Section \ref{sec:amendedDMFT} we relax that assumption. 
Note that in Eq.~(\ref{results:results:5a}), the mean synaptic input depends exponentially on the age of the pattern. 
%This age dependence means that $h_i$ in Eq. (\ref{results:mft:4}) implicitly depends on $s$, which in turn leads to an age dependence of the overlap $m_s$.  
For most recent memories ($\mu\ll K$) the  mean synaptic current is close to its maximum in magnitude $\nu_{\mu}\approx A \eta_i^{\mu} m_{\mu}$, but it decays exponentially for older memories, $\mu \sim O(K)$. 

%Since $h$ depends on $m_s$, the overlap appears in the right and left hand side of Eq. (\ref{results:mft:4}). To self-consistently solve for $m_s$ we return to Eq. (\ref{results:results:1}). 
We next turn to the fluctuations around the mean. The variable $\zeta_i(t)$ is assumed to be a Gaussian random field with mean zero, representing the variability around the mean synaptic inputs  to neuron $i$. Its autocovariance function is given by 
%\begin{equation}
 %\label{results:mft:5}
%\text{Cov}\left(\zeta_i(t),\zeta_i(t+t')\right)=  \frac{A^2\tau}{2} \langle \phi(h(t))\phi(h(t+t')) \rangle_{h,\eta}.
%\end{equation}
%The variable $ \zeta_i (t)$ is a random Gaussian field with zero mean and  auto-covariance  given by
\begin{equation}
 \text{Cov}\left(\zeta_i(t),\zeta_i(t+\tau)\right)= \gamma A^2\kappa \langle \phi(h(t))\phi(h(t+\tau)) \rangle_{h},\label{results:mft:5}
\end{equation}
where 
\begin{equation}
\label{mft:4}
\kappa =\frac{1}{K} \sum_{\mu=1}^{\infty} e^{-\frac{2 \mu}{\tau K}},
\end{equation}
%\Yonatan{Is $\gamma$ necessary? Can it not be absorbed into $A$ WLG?}
%\Ulises{In my opinion is better leve it like that since this parameter shows explicitly how the learning rule affects on the variance of the input currents. For example take $\eta$ binary, log-normal, sparse, etc.}
and $\gamma = \langle \eta^2\rangle ^2.$ In particular, for binary patterns considered here, $\gamma=1$.
In the large $K$ limit, Eq.~(\ref{mft:4}) leads to $\kappa=\tau/2$.

% In the further discussion we will only consider forgetting kernels in which  Eq.~(\ref{mft:4})  converges to a finite number. 
Eqs.~(\ref{results:results:5a},\ref{mft:4}) makes it clear that the number of patterns that can be retrieved is of order $K$. Thus, in the large $N$ limit, we define a continuous age index $s=\mu/K$, such that $s=0$ for recently learned patterns, while $s\to\infty$ for patterns seen in the distant past.% We define the learning rule dependent constant $\gamma =  \langle f^2(\eta) \rangle     \langle g^2(\eta) \rangle$. 

For a pattern with age $s$, the dynamics of the currents can be rewritten as
 \begin{equation}
\label{mft:5}
\frac{{\rm d}h_i}{{\rm d}t}=-h_i+A \eta_i^{s} e^{-\frac{s}{\tau}}m_{s}+A \sqrt{\frac{\gamma \tau}{2} } y_i(t),
\end{equation}
where $y(t)$ is a Gaussian random field with auto-covariance function
\begin{equation}
\label{mft:6}
C_s(t')=\langle y(t)y(t+t')\rangle_y =   \langle \phi(h(t))\phi(h(t+t') \rangle_{h},
 \end{equation}
where the subscript $s$ in the above equation is here to remind us that the autocovariance function depends on the age of the retreived pattern.
By defining the  local currents  $u_i(t)=h_i(t)-A \eta^{s}_i e^{-\frac{s}{\tau}}m_{s}$, Eqs.~(\ref{mft:5},\ref{results:mft:4},\ref{mft:6}) can be re-written as
\begin{eqnarray}
 \dot{u}&=&-u+A \sqrt{\gamma \alpha} y(t)\label{mft:7}\\
m_{\mu}&=&\langle \eta \phi(u(t)+A\eta e^{-\frac{s}{\tau}}m_{s}), \rangle_{u,\eta},\label{mft:8}
\end{eqnarray}
while
\begin{equation}
%\begin{split}
C_s(t')=\langle \phi\left(u(t)+A\eta e^{-\frac{s}{\tau}}m_{s}\right)\phi\left(u(t+t')+A\eta e^{-\frac{s}{\tau}}m_{s}\right) \rangle_{u,\eta}.\label{mft:9}
%\end{split}
\end{equation}

As in \cite{sompolinsky1988chaos,kadmon2015transition} we introduce the synaptic input current auto-covariance function 
\begin{equation}
\label{mft:10}
\Delta_s(t')=\langle u(t)u(t+t')\rangle_{u}.
\end{equation}
where again $s$ reminds us of the age dependence of this autocovariance function. 
Analogously to the derivation in \cite{crisanti2018path,schucker2016functional}, we can derive a self-consistent equation 
for the local-field auto-covariance 
\begin{equation}
\label{mft:11}
\frac{{\rm d}^2 \Delta_s (t')}{{\rm d} \tau^2}=\Delta(t')- \frac{A^2\gamma \tau}{2}  C_s(t').
\end{equation}

% \Yonatan{Our theory predicts a continuum of memory states, each with its own distinct, age-dependent statistics, governed by the self-consistency conditions [Eq. (\ref{mft:9}, \ref{mft:11})]. To simplify the notation we drop the subscript $s$.} 

Using Eq.~(\ref{results:mft:2}), we can evaluate the age-dependent overlap. Eq.~(\ref{results:mft:4}) becomes, 
\begin{equation}
m_{s}=\int D\eta Dx \eta\phi\left(A \left[\sqrt{\Delta_{0s}}x+\eta e^{-\frac{s}{\tau}}m_{s}\right]\right).
\end{equation}
%\Yonatan{The notation here is very confusing. What's the difference between $m_s$ and $m_\mu$? You're using both in the same equation immediately after saying you will drop the age subscript. Given that it is confusing, and given that $\mu$ is used also as the average synaptic input, I suggest keeping the subscript $s$ and using it as much as possible instead of $\mu$.}

We define $\Delta_{0s}$ as the auto-covariance at zero time-lag $t'=0$, i.e., $\Delta_s(0)=\Delta_{0s}$, i.e.,~the variance of the synaptic input currents $h_i$. The auto-covariance in Eq. (\ref{mft:9}) can be written as
\begin{widetext}
\begin{equation}
\label{mft:12}
%\begin{split}
C_s(t')=\int D \eta Dz  \left[\int Dx \phi\left(A\left[\sqrt{\Delta_{0s}-|\Delta_s(t')|} x+\sqrt{|\Delta_s(t')|} z+\eta e^{-\frac{s}{\tau}}m_{s}\right]\right)\right]^2,
%\end{split}
\end{equation}
\end{widetext}
where $D \eta =p_{\eta}(\eta) d\eta$, $Dx = dx e^{-x^2/2}/\sqrt{2 \pi}$, and  $Dz = dz e^{-z^2/2}/\sqrt{2 \pi}$.

We further assume that $\Delta(t')\geq 0$, and re-scale $\Delta(t')$, $\Delta(t') \to A^2 \Delta (t')$.  As in  \cite{sompolinsky1988chaos},
Eq.~(\ref{mft:11}) can be re-written in terms of a potential 
\begin{equation}
\label{mft:13}
\frac{{\rm d}^2 \Delta_s }{{\rm d} t'^2}=-\frac{\partial V(\Delta_s,\Delta_{0s})}{\partial \Delta_s },
\end{equation}

by defining 
\begin{widetext}
\begin{equation}
\label{mft:14}
%\begin{split}
V(\Delta_s,\Delta_{0s})=-\frac{\Delta_s^2}{2}+\frac{\kappa \gamma}{A^2}\int D \eta Dz \left[\int Dx \Phi\left(A\left [ \sqrt{\Delta_{0s}-|\Delta_s|} x+\sqrt{|\Delta_s|} z+\eta e^{-\frac{s}{\tau}}m_{s}\right]\right)\right]^2,
%\end{split}
\end{equation}
\end{widetext}
where $\Phi(x)=\int_{0}^xdr \phi(r)$.
%Notice that  analogous to $m_{s}$, the auto-covariance of the local fields $\Delta_0$ also depends on the age of the retrieved memory $s$, however we choose to not make explicit this dependency in order to simplify the notation.

The dynamics in Eqs.~(\ref{mft:13}, \ref{mft:14}) corresponds to equation of particle in a Newtonian potential $V(\Delta_s, \Delta_{0s})$ that depends parametrically on $\Delta_{0s}$. As shown by \cite{sompolinsky1988chaos}, the motion of the particle should be such that $\frac{d \Delta_{s}(0)}{dt'}=0$ and $ V(\Delta_{0s}, \Delta_{0s}) = V(\Delta_{s}(\infty), \Delta_{0s})$. Therefore, the network dynamics in Eq.~(\ref{mft:5}) is described by three {\it order parameters}: the overlap $m_s$; the synaptic input current variance $\Delta_{0s}$; and $\Delta_{1s}$, which is the long time limit of the auto-covariance of the synaptic input currents $\Delta_{1s}=\lim_{t'\to \infty}\Delta_s(t')$. These can be computed self-consistently, as shown below. 
%We call the solutions to these equations the DMFT solution to the network dynamics.  
 
 As we will see below, there are four qualitatively different types of solutions to the DMF equations. Solutions with $m_s=0$ correspond to background states (i.e.,~states that are uncorrelated with all stored memories), while solutions with $m_s>0$ correspond to memory (or retrieval) states, since there exists a finite overlap between network state and the corresponding stored pattern. Both types of solutions can be either fixed-point attractors of the dynamics (i.e.,~with autocovariance functions that are constant in time, i.e.,~ $\Delta_s(t)=\Delta_{0s}$ at all times) or chaotic attractors, with an autocovariance function that depends on time. We first focus our attention on the transitions to chaos, and then study the capacity of the system (largest age at which retrieval states still exist).
%By making the assumption that memory states are static fixed-point attractors, and therefore there are no temporal fluctuations in the stationary memory state, then the temporal variability in our theory is quenched $y_i(t)=y_i$ (see Eq.~(\ref{mft:6})). This implies that $\Delta_1 = \Delta_0$, and the dynamics are described by two simpler equations instead of the three in the DMFT (see Eqs.~(\ref{bif:1}, \ref{bif:2})). We call the solutions to these equations the static mean field theory (SMFT) solutions to the network dynamics. The SMFT and DMFT solutions will coincide only when retrieval states do not vary in time. 
 
%Importantly, our theory provides quantitative predictions for the transitions to qualitatively different types of dynamics depending on the network parameters (the memory age $s$;  the update strength $A$; and the forgetting time scale $\tau$).  We assess the accuracy of these predictions by comparing our theory with simulations of large and sparse networks. 

\subsection{Transitions to chaos\label{sec:transitionchaos}}

%In this section we determine the location in parameter space 
%where fixed-point attractors transition to chaotic attractors. 
% We distinguish two qualitatively different attractor states depending on the overlap with the stored memory: 1) states with order one overlap (i.e., $m_{\mu}\sim \mathcal{O}(1)$) we call memory states; 2) states with negligible overlaps (i.e., $m_{\mu}\ll 1\quad \forall \mu$) we call the background state.

%  Transitions to chaos of memory states
%\subsubsection{Transition to chaos of fixed-point attractor memory states \label{sec:chaos:chaosfixed}}

In fixed-point attractors there are no temporal fluctuations in the input currents.  The auto-covariance of the local fields in 
 Eq.~(\ref{mft:10})  is then equal to the variance of the local currents at all times  (i.e., $\Delta_s(\tau)=\Delta_{0s}$), which leads to
\begin{eqnarray}
m_{s}&=&\int D\eta Dx \eta\phi\left(A \left[\sqrt{\Delta_{0s}}x+\eta e^{-\frac{s}{\tau}}m_{s}\right]\right)\label{bif:1}\\
\Delta_{0s}&=&\gamma \kappa \int D\eta Dx \phi^2\left(A \left[\sqrt{\Delta_{0s}}x + \eta e^{-\frac{s}{\tau}}m_{s}\right]\right).\label{bif:2}
\end{eqnarray}
%\Yonatan{Given that we must show these equations here, why do we introduce the distinction between SMFT and DMFT in the previous section? Wouldn't it be clearer to say: "To study the transition to chaos, we assume that there are no temporal fluctuations, which gives Eqs. 23, 24, instead of 3 equations for the DMFT. When the solutions to these equaitons are unstable/don't exist, then ...}

The above equations give the overlap in a retrieval fixed point attractor state of a memory of age $s$. However, these fixed points can destabilize and become chaotic, leading to chaotic retrieval states. Importantly, as we will see below in this model the dynamical properties of the attractors (i.e., fixed-point or chaotic) strongly depend on the age of the patterns. Eqs.~(\ref{bif:1},\ref{bif:2}) have solutions in parameter space even beyond the transition to chaos, when the assumption of fixed-point attractor memory states with no temporal fluctuations in the input currents is no longer valid. However, as expected, their predictions depart from network simulations (See Fig.~\ref{fig:1} dashed line; Fig.~\ref{fig:2} A and B red lines). We refer to the solutions of these equations the static solutions to DMFT (SMFT). 
 
Analogous to  \cite{sompolinsky1988chaos}, to find the  transition to chaos of memory states, it is necessary to find the point in parameter space where the static solution $\Delta(t')=\Delta_0$ becomes unstable. At this point  the auto-covariance of the local-field $\Delta(t')$ transitions from stationary to time-dependent. Since the dependence on time of the auto-covariance of the local fields is ruled by the Newtonian equation for the position $\Delta$ at ``time'' $t'$ of a particle subject to a potential energy (Eq.~(\ref{mft:11})), finding the transition point is equivalent to finding the critical point  $\Delta_0^{\text{chaos}}$ where the potential in Eq.~(\ref{mft:14}) changes its concavity. After this point, solutions for the auto-covariance of the local field starting at $\Delta_0$  relax to $\lim_{t'\to \infty}\Delta(t') \equiv \Delta_1$. The transition point is given by
 \begin{equation}
\label{bif:3}
A^2 \gamma \kappa \int D \eta Dz  \left\{\phi^{'}\left(A\left [\sqrt{\Delta_{0s}} z+ \eta e^{-\frac{s}{\tau}}m_{s}\right]\right)\right\}^2=1.
\end{equation}
 Equation~(\ref{bif:3}) in addition to Eqs.~(\ref{bif:1},\ref{bif:2}) describe the curve in the parameter space that separates fixed-point from chaotic memory states. 

%TRANSITION CHAOS BACKGROUND
%\subsubsection{Transition to chaos of the background state ($m=0$)}

In the background state, all the overlaps with the stored memories are zero, i.e.,  $m_s = 0$.  The critical line in the space of parameters 
for its transition to chaos is thus given by the following set of equations
\begin{eqnarray}
&&A^2 \gamma \kappa\int Dz  \left\{\phi^{'}\left(A \sqrt{\Delta_0} z\right)\right\}^2=1  \label{bif:4} \\
&&\Delta_0=\gamma \kappa \int Dz \phi^2\left(A \sqrt{\Delta_0}z\right). \label{bif:5}
\end{eqnarray}

%CAPACITY
%HERE
 \subsection{Capacity\label{sec:capacity}}

As there exist two types of memory states, there are two possible ways to compute the capacity: the maximum age at which fixed-point or chaotic retrieval states cease to exist, respectively. However, as we will see later, in a range of parameters of the system the physically relevant capacity is the one computed from chaotic retrieval states, since fixed point memory states destabilize before reaching capacity for those parameters.
%The capacity of fixed-point  attractor states is defined by the largest age $s$ that leads to a non-zero overlap with the corresponding pattern, or equivalently the smallest $s$ for which $m_{s}=0$ is the  only solution of Eqs.~(\ref{bif:1},\ref{bif:2}). In this scenario the underlying  assumption is that memory states are fixed-points. However,  in this model chaotic memory states may undergo a transition to zero overlap with the stored pattern. 
In chaotic attractors, the potential defined in Eq.~(\ref{mft:14}) is not convex (i.e., $\partial^2 V/\partial \Delta^2>0$), and the auto-covariance of the local currents in Eq.~(\ref{mft:10}) is time dependent. 
 Additionally,  as shown by \cite{sompolinsky1988chaos, kadmon2015transition}, a chaotic solution is characterized by an aperiodic, decreasing
  potential. This corresponds to the condition $\lim_{t'\to \infty}V(\Delta_s(t'))=V(\Delta_{0s})$, which is equivalent to
 \begin{widetext}
 \begin{equation}
\label{overlap:chaos:1}
%\begin{split}
-\frac{\Delta_{0s}^2}{2}+\frac{\kappa \gamma}{A^2}  \int D\eta D x  \Phi^2\left(A\left[\sqrt{\Delta_{0s}} x+\eta e^{-\frac{s}{\tau}} m_s\right]\right)=-\frac{\Delta_{1s}^2}{2}+
\frac{\kappa \gamma}{A^2}\int D \eta  Dz \left[\int Dx \Phi\left(A\left [ \sqrt{\Delta_{0s}-|\Delta_{1s}|} x+\sqrt{|\Delta_{1s}|} z+\eta e^{-\frac{s}{\tau}} m_s\right]\right)\right]^2,
%\end{split}
\end{equation}
\end{widetext}
 where  $\Delta_{1s}$ corresponds to the large time limit of $\Delta_s(t)$, $\Delta_{s}(t)\xrightarrow{t \to \infty}\Delta_{1s}$. Therefore,   $\left.\frac{\partial V}{\partial \Delta}\right|_{\Delta = \Delta_{1s}}=0$, which  is equivalent to 
 \begin{widetext}
  \begin{equation}
\label{overlap:chaos:2}
\Delta_{1s} = \kappa \gamma   \int D \eta  Dz \left[\int Dx \phi\left(A\left [ \sqrt{\Delta_{0s}-|\Delta_{1s}|} x+s\sqrt{|\Delta_{1s}|} z+\eta e^{-\frac{s}{\tau}} m_s\right]\right) \right]^2.
\end{equation}
\end{widetext}
In Eqs.~(\ref{overlap:chaos:1},\ref{overlap:chaos:2}), the overlap $m_s$ is given by Eq.~(\ref{bif:1}). Thus, the three order parameters $m_s$, $\Delta_{0s}$ and $\Delta_{1s}$ can be obtained by solving numerically Eqs.~(\ref{bif:1},\ref{overlap:chaos:1},\ref{overlap:chaos:2}). For fixed-point attractor memory states, there are no temporal fluctuations in the input currents and therefore   $\Delta_{1s}=\Delta_{0s}$, recovering from  Eqs.~(\ref{bif:1},\ref{overlap:chaos:1},\ref{overlap:chaos:2}) the static solutions in Eq.~(\ref{bif:1},\ref{bif:2}). Beyond the transition to chaos, the overlap curve for chaotic attractors is given by Eqs.~(\ref{bif:1},\ref{overlap:chaos:1},\ref{overlap:chaos:2}). 
%In this model, we can compute analytically the capacity curve in parameter space $(s, \tau, A)$ beyond these memeory states cannot be retrieved. , which is the par 
The capacity of this network corresponds to the older memory age $s_c$ beyond which no older memory can be retrieved. At capacity, the corresponding memory state has zero overlap $m_{s_c}=0$. Since the transfer function in this model ($\phi(x) = \tanh(x)$) is odd,  at capacity we have that $\Delta_{1s_c}=0$ (see Eq.~(\ref{overlap:chaos:2})). The value of $\Delta_{0s_c}$ is then obtained by solving Eq.~(\ref{overlap:chaos:1}), and is given by
\begin{equation}
\label{TT:10}
\tau =\frac{(A \Delta_{0s^c})^2}{\left[ \int D x  \Phi^2\left(A\left[\sqrt{\Delta_{0s^c}} x\right]\right)-\left(\int  Dx \Phi\left(A\left [ \sqrt{\Delta_{0s^c}} x \right]\right)\right)^2 \right]},
\end{equation}
with
%   I am not sure if $\phi(x)=-\phi(-x)$ is enough for derivating Eq.~(\ref{overlap:static:1}) with respect of $m$  and evaluating at $m=0$ for obtaining $\Delta_0^c$, but definitely with $\phi(x)=\tanh(x)$ works, 
\begin{equation}
\label{TT:11}
A e^{-\frac{s^c}{\tau}} \int Dx \phi^{'}\left(A \left[\sqrt{\Delta_{0s^c}}x\right]\right)=1.
\end{equation}

Eqs.~(\ref{TT:10},\ref{TT:11}) provide the capacity curve in parameter space ($s_c,\tau, A$) (see the curve that separates the green and the gray region in Fig.~\ref{fig:4}). Notice that Eq.~(\ref{TT:11}) is obtained by derivating Eq.~(\ref{bif:1}) with respect of the overlap and evaluating the resulting equation at capacity (i.e., $m_{s_c}=0$), considering the overlap changes continuously with the memory age (see Fig.~\ref{fig:5}A). 
 
\subsection{Stability of memory states\label{sec:amendedDMFT}}

In this section, we study the stability of a memory state of a given age $s$ with respect to perturbations in neuronal activity in the direction of another, more recent, memory states. Thus, we assume that the network state is correlated with two memories: the currently retrieved one, of age $s$, but also a recent memory ($s=0$). For simplicity we focus on the case of binary patterns. The aim of this calculation is to derive dynamical equations for the transient dynamics of the overlaps $m_0$ and $m_s$, similarly to a recent calculation for recurrent networks storing sequences of activity \cite{gillett2020characteristics}. Then, we can use these equations to study numerically the stability of the memory state corresponding to the memory $s$, i.e.,  $m_s\sim \mathcal{O}(1)$ and $m_0\ll 1$. 

Similarly to Eq.~(\ref{mft:5}), the dynamics of the currents are approximated by a time dependent Gaussian random field 
\begin{equation}
\label{amendeddmft:1}
\frac{{\rm d}h_i}{{\rm d}t}=-h_i+A (\eta_i^{0}m_0+\eta_i^{s} e^{-\frac{s}{\tau}}m_{s})+A \sqrt{\frac{ \tau}{2} } y_i(t).
\end{equation}
At any time, the synaptic input current $h_i(t)$ can be described by a Gaussian random variable 
 \begin{equation}
\label{amendeddmft:2}
h(t)=A (\eta_i^{0}m_0+\eta_i^{s} e^{-\frac{s}{\tau}}m_{s})+\sqrt{\frac{ \tau\Delta_{0s}(t)}{2}  }x,
\end{equation}
where $x$ is Gaussian random variable with mean 0 and variance 1, and the variance $\Delta_{0s}$ is now time dependent. 
Using the above in Eq.~(\ref{results:model:1}) we obtain dynamical equations for the overlaps 
\begin{widetext}
 \begin{eqnarray}
 \frac{{\rm d}m_0}{{\rm d}t}&=&-m_0+ \frac{1}{2}\int Dx\left[\phi\left(A\left[m_0+e^{-\frac{s}{\tau}}m_{s} + \sqrt{\frac{ \tau\Delta_{0s} }{2} }x\right]\right)+\phi\left(A\left[m_0-e^{-\frac{s}{\tau}}m_{s} + \sqrt{\frac{ \tau\Delta_{0s} }{2} }x\right]\right)\right]\label{amendeddmft:5}\\
  \frac{{\rm d}m_s}{{\rm d}t}&=&-m_s+\frac{1}{2}\int Dx\left[\phi\left(A\left[m_0+e^{-\frac{s}{\tau}}m_{s} + \sqrt{\frac{ \tau\Delta_{0s} }{2} }x\right]\right)+\phi\left(A\left[e^{-\frac{s}{\tau}}m_{s}-m_0 + \sqrt{\frac{ \tau\Delta_{0s} }{2} }x\right]\right)\right].\label{amendeddmft:6}
 \end{eqnarray}
  \end{widetext}
The dynamics of the variance $\Delta_{0s}$ and the two-time autocorrelation function can be derived by a system of  integro-differential equations (see section 2.2 in the supplementary information of \cite{gillett2020characteristics} for a derivation in a closely related model).
Here, we choose a simpler approach and approximate the dynamics of $\Delta_{0s}$ and $\Delta_{1s}$ 
%in Fig.~\ref{fig:5} A and B and Fig.~\ref{fig:6}A, 
by a system of ODEs (see Appendix \ref{appendix:relaxational} for more details) whose steady states are given by \begin{widetext}
 \begin{equation}
\label{amendeddmft:8}
\begin{split}
&-\Delta_{0s}^2+\frac{\tau }{2A^2}  \int  D x \left\{  \Phi^2\left(A\left[\sqrt{\Delta_{0s}} x+m_0+e^{-\frac{s}{\tau}}m_{s}\right]\right)+\Phi^2\left(A\left[\sqrt{\Delta_{0s}} x+e^{-\frac{s}{\tau}}m_{s}-m_0\right]\right)\right\}=-\Delta_{1s}^2+\\
&\frac{\tau}{ 2A^2}\int   Dz\left\{ \left[\int Dx \Phi\left(A\left [ \sqrt{\Delta_{0s}-|\Delta_{1s}|} x+\sqrt{|\Delta_1|} z +m_0+e^{-\frac{s}{\tau}}m_{s}\right]\right)\right]^2+\right.\\
&\left.\left[\int Dx \Phi\left(A\left [ \sqrt{\Delta_{0s}-|\Delta_{1s}|} x+\sqrt{|\Delta_{1s}|} z+e^{-\frac{s}{\tau}}m_{s}-m_0\right]\right)\right]^2\right\}
\end{split}
\end{equation}
\end{widetext}
and 
 \begin{widetext}
  \begin{equation}
  \begin{split}
\label{amendeddmft:7}
\Delta_{1s} = \frac{\tau}{4}  \int Dz\left\{ \left[\int Dx \phi\left(A\left [ \sqrt{\Delta_{0s}-|\Delta_{1s}|} x+\sqrt{|\Delta_{1s}|} z+m_0+e^{-\frac{s}{\tau}}m_{s}\right]\right) \right]^2+\right. \\
\left. \left[\int Dx \phi\left(A\left [ \sqrt{\Delta_{0s}-|\Delta_{1s}|} x+\sqrt{|\Delta_{1s}|} z+e^{-\frac{s}{\tau}}m_{s}-m_0\right]\right) \right]^2\right\}.
\end{split}
\end{equation}
\end{widetext}

% similarly to Eqs.~(\ref{overlap:chaos:1},\ref{overlap:chaos:2}) for $\Delta_{0s}$ and $\Delta_{1s}$. 
  
Neglecting the dynamics of the $\Delta$ parameters, and expanding the overlap equations Eqs.~(\ref{amendeddmft:5},\ref{amendeddmft:6}) around the memory state $m_s$ and $m_0=0$ we obtain following instability line 
 \begin{equation}
     \label{amendeddmft:6a}
     \int Dx\phi^{'}\left(A\left[e^{-\frac{s}{\tau}}m_{s} + \sqrt{\frac{ \tau\Delta_{0s} }{2} }x\right]\right)=\frac{1}{A}.
\end{equation}
 
Our network simulations show that this approximation matches well the network dynamics (see Fig.~\ref{fig:5} and \ref{fig:6}) suggesting that the observed memory state instability is mainly governed by the overlap dynamics in Eqs.~(\ref{amendeddmft:5},\ref{amendeddmft:6}).  
%Reproducing the network transient activity is beyond the scope of this paper, here we focus on the stability of the different solutions. %\Nicolas{Explain what you are doing exactly with with Deltas - you say constant but the equations depend on overlaps who are themselves time-dependent}

%The above Eqs.~(\ref{amendeddmft:5},\ref{amendeddmft:6}) describe the transient dynamics of the overlaps  with the most recent $m_0$ and the older $m_s$ memories for any initial condition. Here we approximate the covariance $\Delta_{0s}$ and the autocovariance  $\Delta_s(t')$ of the synaptic input currents as constant. However, for any initial condition they also change in time, and its transient dynamics is described by a system of  integro-differential equation that it has been calculated for similar networks (see section 2.2 in the supplementary information of \cite{gillett2020characteristics}). Reproducing the network transient activity is beyond the scope of this paper. Here we focus on the stability of the different solutions, so we  approximated $\Delta_{0s}$ and $\Delta_s('t)$ as constant in time. In Section \ref{sec:netowork_forgetting} we show that this approximation is sufficient to make an accurate  prediction for the memory age $s^*$ above which networks jump form an older to the most recent memory state.

%T.T model

% sparsely connected Hopfield model 
\section{Dynamics in the absence of forgetting\label{sec:tirozzitsodyks}}

We start by analyzing the dynamics of this network in the limit where there is no forgetting ($\tau\to \infty$). In this scenario, we need to consider a finite number of presented patterns $p$, since the variance of synaptic weights diverges in the limit $p\rightarrow\infty$, and the synaptic connectivity can be written as
\begin{equation}
\label{results:slow:1}
J_{ij}=\frac{A c_{ij}}{Nc} \sum_{\mu=1}^{p} \eta^{\mu}_i\eta^{\mu}_j.
\end{equation}

\begin{figure*}%[htbp]
\includegraphics[width=1.\textwidth]{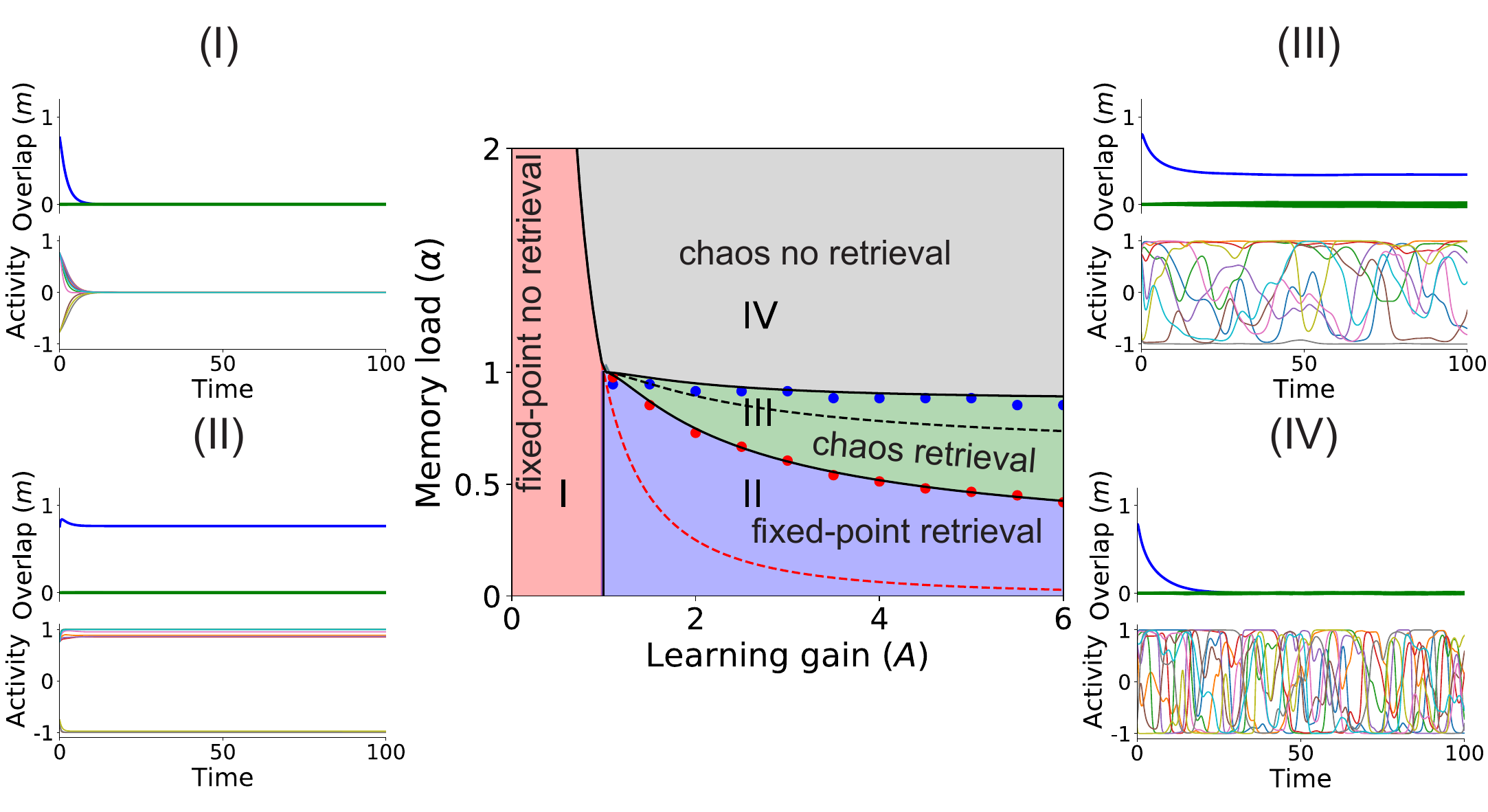}
\caption{Center: Bifurcation diagram for the TT model in the parameter space spanned by memory load ($\alpha$) vs the learing gain ($A$). Full lines: Boundaries of the four qualitatively different regions calculated using DMFT. Red circles: location of the transition to chaos of retrieval states, using simulations of networks of $N=10^7$ units and $K= 2 \log(N)$ connections per neuron. Blue circles: storage capacity, computed using simulations. Dashed red line: Transition to chaos of the background state. Black dashed line: Line on which (unstable) fixed point retrieval states disappear. Surrounding panels, labelled I-IV, show representative numerical simulations in the four qualitatively different regions. In each simulation the network is initialized close to one of the stored memories. Each plots show the overlaps of network state with this memory (blue), as well as other memories (green); and the activity of 10 randomly selected neurons, as a function of time. The networks parameters $A$ and $\alpha$ are indicated by the location of the bottom left corner of the corresponding roman numbers.  The network parameters for the surrounding panels are $N=10^6$ and $K= 2 \log(N)$. The parameter values for  $A$ and $\alpha$ for panels I-IV are respectively $A=0.5, 2.5, 2.5, 2.5$ and $\alpha = 11/K, 11/K, 21/K, 33/K$.}
\label{fig:1}
\end{figure*}

\begin{figure*}%[htbp]
\includegraphics[width=1.\textwidth]{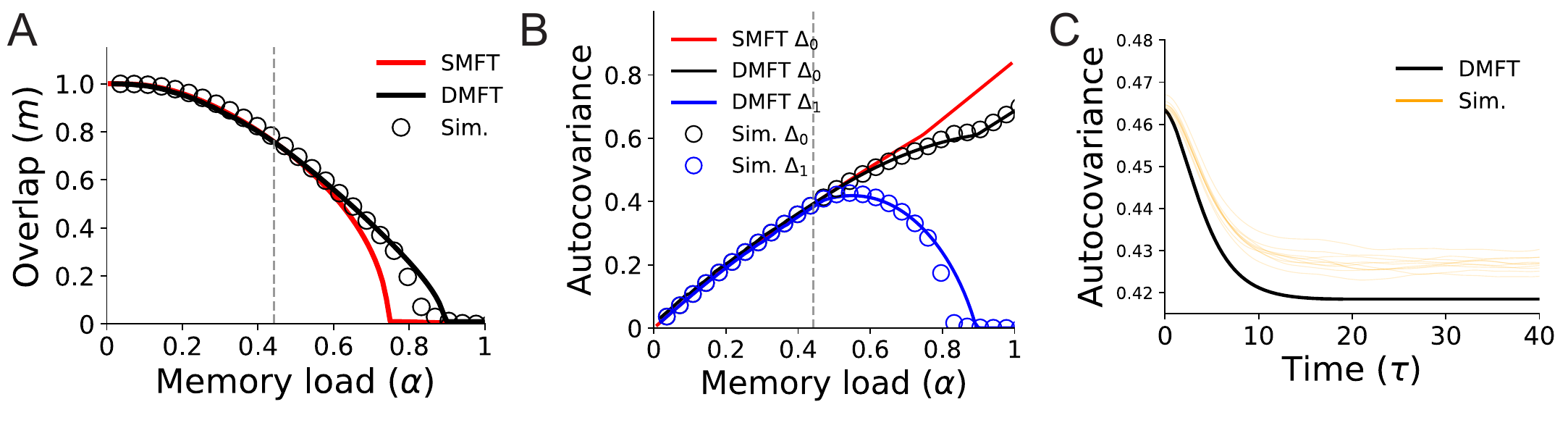}
\caption{(A) Overlap vs memory load. Circles: Overlaps computed from
network simulations (average over 10 realizations). Dashed vertical line: Transition to chaos. Black lines overlaps computed from DMFT. Red lines:
Overlaps in static solutions to DMFT (SMFT) which are unstable beyond the transition to chaos. (B) Parameters characterizing the autocovariance function $\Delta_{0s}$ and $\Delta_{1s}$. As in A, full lines represent solutions from MFT, while circles represent simulation results. (C) Autocovariance function for $\alpha=0.5429$ (i.e., $\alpha =15/K$) (Black: DMFT; Yellow: 10 realizations of network simulations).
Other parameters, $N=10^6$, $K= 2 \log(N)$, and  $A = 5.5$.}
\label{fig:2}
\end{figure*}

\begin{figure}%[htbp]
%\begin{adjustwidth}{-2.3in}{0cm}
\includegraphics[width=.5\textwidth]{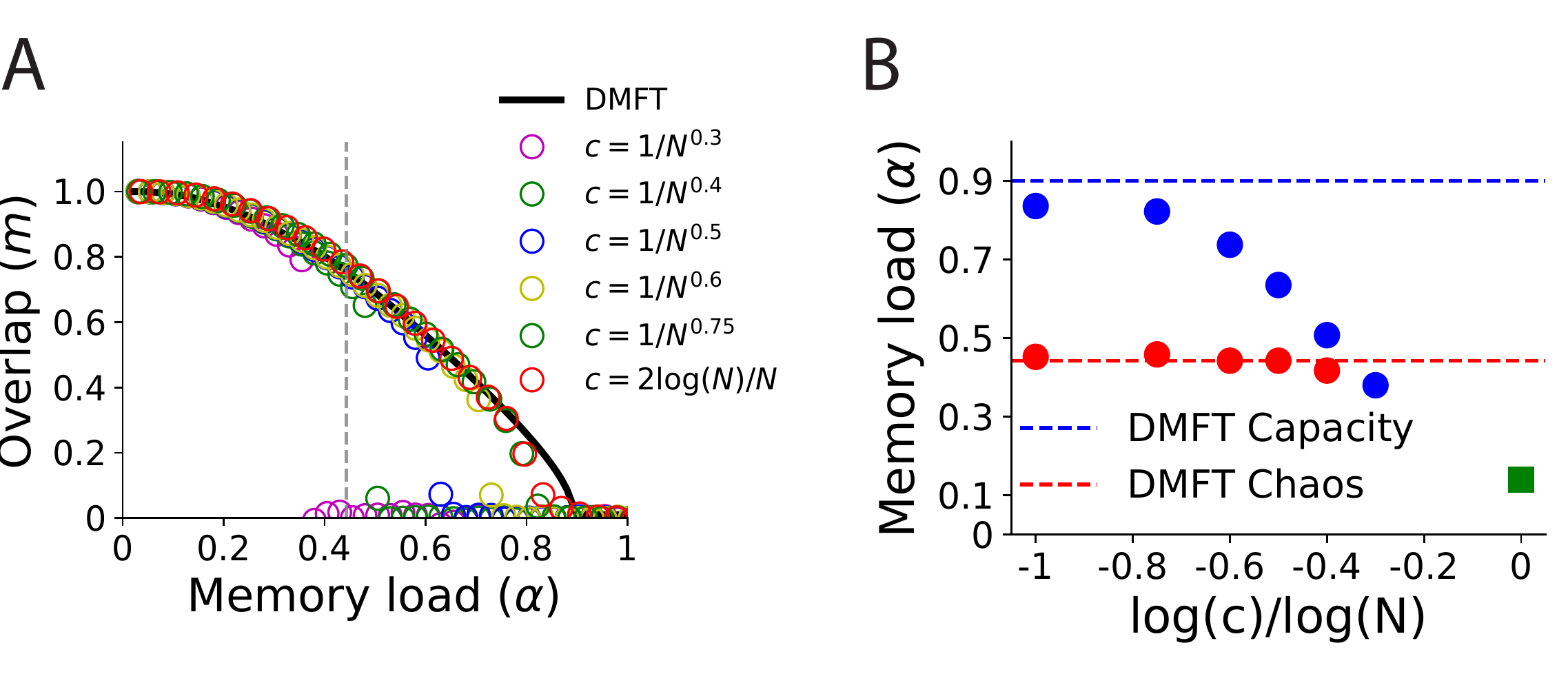}
\caption{Effect of connectivity sparseness on the dynamics of the TT model. (A) Overlap vs memory load, for different scalings of connection probability with network size $N$.  Circles: Overlap averaged time  after  transients  and  over  10  realizations.   Horizontal dashed  line: transition to chaos.  Black line: Overlap computed from DMFT. (B) Capacity vs $\log(c)/\log(N)$ (blue circles) and  transition to chaos vs $\log(c)/\log(N)$ (red circles). The capacity is estimated from network simulations as the minimal memory load for which the overlap is smaller than 0.1. Dashed lines: DMFT red line corresponds to the capacity predicted by the DMFT. Notice that for $\lim_{N\to \infty} \log(\log(2\log(N))/N)/\log(N) = -1$. Green square: capacity in the fully connected case \cite{kuhn1991statistical} which is equal to the capacity of the Hopfield model (0.14). Networks sizes corresponding to the scalings $c =1/{N^{0.3}},1/{N^{0.4}}, 1/{N^{0.5}}, 1/{N^{0.6}}, 1/{N^{0.75}}, 2 \log(N)/N$ are $N = 10^5, 2.5 \cdot 10^5, 5\cdot 10^5,  10^6, 10^6, 10^6$ respectively. The learning gain is $A=5.5$. }
\label{fig:3}
%\end{adjustwidth}
\end{figure}

Our network model therefore coincides with the model studied by  Tirozzi and Tsodyks \cite{tirozzi1991chaos} (referred to as TT model in the following) who provided an analytical description of the network qualitative different behaviors in the sparse coding limit. In the fully connected case, $c_{ij}=1$, it coincides with the model introduced by Hopfield \cite{hopfield1984neurons} and studied analytically by Kuhn et al \cite{kuhn1991statistical}. In this limit, the sum on the r.h.s.~ of Eq.~(\ref{mft:4}) becomes a sum over $p$ patterns, and $\kappa$ becomes equal to the memory load $\alpha:=\kappa=p/K$, i.e.,~ the number of stored patterns divided by the average number of connections. Here we recapitulate the analytical results provided by Tirozzi and Tsodyks and complement them with simulations of large networks, to investigate how well the theory matches results of such simulations at various degrees of sparsity (see Figs.\ref{fig:1}-\ref{fig:3}).

In the TT model, the network dynamics depends on two parameters,
the memory load $\alpha = p/K$, i.e.,~the number of stored patterns scaled by the average number of connections, and $A$, the strength by which the patterns are imprinted in the connectivity when learned. Figure \ref{fig:1} shows the network bifurcation diagram derived from the DMFT delineating the four different dynamical regimes in parameter space, in which the four qualitatively different types of states described in Section \ref{sec:DMFT} exist: I) fixed point background state. II) fixed point memory states (i.e., states with a finite overlap $m\sim \mathcal{O}(1)$ with the stored patterns); III) chaotic memory states (i.e., with a finite overlap $m\sim \mathcal{O}(1)$ with stored patterns); IV) chaotic background state.  For small values of the update strength $A$, the network is weakly coupled, and only the background state, in which the firing rates of all neurons are equal to zero, exists. In this regime, memory retrieval is not possible, and the overlap decays to zero when the network is initialized with any of the stored patterns (see red region I in Fig.~\ref{fig:1}). For larger values of $A$ and small memory load, any of the stored patterns can be retrieved as fixed-point attractor states. When the network is initialized close to one of the stored patterns, the network goes to a fixed-point that is correlated with that pattern (see region II in blue in Fig.~\ref{fig:1}). Within this parameter region, the red dashed line in  Fig.~\ref{fig:1} indicates the transition to chaos of the background state. Below the red dashed line, for small memory load, the background state is a fixed-point while for larger memory loads it is a chaotic state. Larger memory loads lead a transition to chaos of the memory states (green region III in Fig~\ref{fig:1}). In this regime, the network dynamics are  chaotic, but it retains a finite overlap with the retrieved memory. Therefore, in spite of chaotic fluctuations, the population activity is stably correlated with the retrieved pattern.  Finally, if the memory load further increases none of the stored patterns can be retrieved and all memories are forgotten. This is a well known phenomenon in attractor neuronal networks called {\it catastrophic forgetting}. The value of the memory load when this happens is called the memory capacity. Beyond the memory capacity, the only stable attractor is the chaotic background state (gray region in Fig~\ref{fig:1}). The dashed black line in Fig.~1 corresponds to the memory load at which (unstable) fixed point retrieval states disappear. Interestingly,
this line is below the line at which chaotic retrieval states disappear (compare dashed line with solid line between the green and gray region in Fig.~\ref{fig:1}). Thus, chaotic fluctuations allow the network to increase the storage capacity of the network.

Fig.~\ref{fig:2} compares the results from DMFT with numerical simulations of a very large and sparse network ($N=10^6$, $K=2 \log(N)$). For these parameters the DMFT is in excellent agreement with numerical simulations, as shown by comparing the overlaps in memory states (Fig.~\ref{fig:2}A), parameters of the autocovariance (Fig.~\ref{fig:2}B), and the full autocovariance function (Fig.~\ref{fig:2}C).
We next numerically investigate the validity of our theory for different scalings of the sparsity with the network size. In networks in which the average number of connections scale logarithmicaly with the network size (i.e., $K\sim \log(N)$ and $c\sim \log(N)/N$), the storage capacity computed from network simulations is very close to the to the capacity predicted by the DMFT (see Fig.~\ref{fig:3} A and B). Denser scalings gradually decrease the capacity of the network, and consequently the DMFT capacity predictions become less accurate (see Fig.~\ref{fig:3}A and B). The transition to chaos is well predicted by DMFT in networks in which the connection probability $c=1/N^a$ with $a\ge 0.4$. For less sparse connectivity, retrieval states disappear before the transition to chaos is reached, and therefore chaotic retrieval states non longer exist. For the sake of completeness, we also plot in Fig.~\ref{fig:3}B the capacity in fully connected networks, that was obtained analytically by Kuhn et al \cite{kuhn1991statistical}.
%all scalings for which the capacity is larger than the transition to chaos match very well the DMFT predictions (see Fig.~\ref{fig:3}C). For networks with denser sparsity scalings in which the capacity is close to the transition to chaos, e.g.,  $c=1/N^{0.4}$ (i.e., $K = N^{0.6}$) (see Fig.~\ref{fig:3}A, green circles) the match is less good. For even denser sparsity scalings the capacities are smaller than the transition to chaos and chaotic memory states are lost. 

\section{Dynamics of the network with forgetting \label{sec:netowork_forgetting}}

\begin{figure*}[htbp]
\includegraphics[width=1.\textwidth]{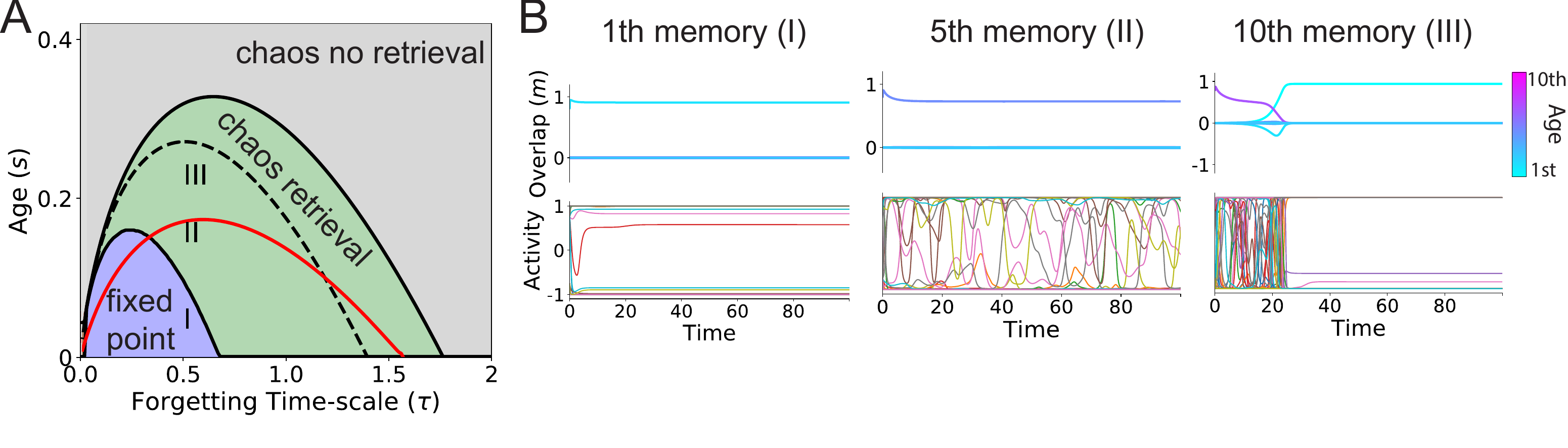}
\caption[Bifurcation diagram for the network with forgetting]{(A) Bifurcation diagram for the network with forgetting, for $A=10$. Solid lines: transition to chaos, and maximal age at which retrieval states exist, calculated using the DMFT. Dashed line: Maximal age at which static retrieval solutions exist (SMFT). Red line: Maximal age at which retrieval states are stable. (B) Retrieval dynamics of memories of age 1, 5 and 10 for the same realization of the connectivity matrix ($N=10^6$, $K= 2 \log(N)$, $A=10$, $\tau=0.5$). These ages correspond to $s= 1/K, 5/K, 10/K$ respectively, see I-III in panel A. } 
%\Nicolas{How does this phase diagram change with $A$? Right panel seems to indicate background state unstable - is this typical?} \Ulises{Q1: Do you suggest another plot for A vs tau? Q2: Yes, this is typical. The background state seems to be unstable.}}
\label{fig:4}
\end{figure*}

We now turn to a network where connectivity is learned using an online Hebbian synaptic plasticity rule, Eq.~(\ref{results:model:2}). In this rule, recent patterns are imprinted in network connectivity using the covariance between pre and post-synaptic activity, while older patterns are forgotten with a decay time-scale $\tau = -1/(K\log(\rho))$. We assume that an infinite number of patterns has been presented to the network, leading to a connectivity matrix given by Eq.~(\ref{results:model:3}).
%Under this assumption the number of patterns is much larger than the average number of connections (i.e., $p/K\to \infty$), which implies that the continuous index $s = \mu/K$ representing the pattern age runs from zero to infinity ($s\in[0,\infty)$). The mean of the synaptic weights in Eq.~(\ref{results:forgetfull:1}) is zero and its variance is bounded due to the exponential forgetting kernel. Therefore, despite the connectivity is the sum of an infinite number of learned patterns, the synaptic weights take finite values. 

In this network, the retrieval dynamics of each memory state depends on its age, and there is a continuum of statistically distinguishable memory states parameterized by the memory age (see section \ref{sec:DMFT}). In a given memory state $s$, the mean input current depends exponentially on the age of the pattern (see Eq.~(\ref{mft:5})). Due to the implicit dependence of the input current autocovariance function on the mean input current (see Eq.~(\ref{mft:12})),  both the mean and the autocovariance of the synaptic input currents $h_i$ vary with the age of the memory. For sufficiently large $A$, we identify using DMFT three different regions depending on the age $s$ and the memory time scale (Fig.~\ref{fig:4}): A region in which retrieval states are fixed point attractors, another region in which retrieval states are chaotic attractors, and finally a region in which no retrieval is possible. Depending on the forgetting time scale, different scenarios as possible as memories age. For short forgetting time scales ($\tau<0.34$ in Fig.~\ref{fig:4}A), most recent memories are retrieved as fixed point attractors (blue region in Fig.~\ref{fig:4}A, and Fig.~\ref{fig:4}B-I). As memories age, they become unstable (red line in Fig.~\ref{fig:4}), and the network typically retrieves one of the most recently learned patterns instead.  For larger forgetting time scales (0.34$<\tau <$0.68 in Fig.~\ref{fig:4}), most recent memories are also retrieved as fixed point attractors. However, unlike for shorter forgetting time scales, fixed point attractor states become chaotic retrieval states after a certain age (above the line separating blue and green regions in Fig.~\ref{fig:4} see an example in Fig.~\ref{fig:4}B-II). As memories further age, there is a second transition line where the chaotic attractor becomes unstable and the network retrieves one of the most recent memories instead (see Fig.~\ref{fig:4}B-III). For even larger forgetting time scales (0.68$<\tau<$1.58), fixed point attractor states no longer exist, and even the most recent memories are retrieved as chaotic attractors. Finally, when $\tau>1.58$ the network is no longer able to retrieve any memory. Note that the maximal capacity (defined as the maximal age at which memories can be retrieved) is given by the red line in Fig.~\ref{fig:4}. Above this line, retrieval states still exist in a finite region of parameter space, but they are unstable and the network retrieves more recent patterns instead. Note also that the capacity is optimized at an intermediate value of $\tau\sim 0.6$ for which both types of retrieval states coexist in the network phase space (fixed point attractors for recent memories, chaotic attractors for older ones). For this value of $\tau$, the capacity is about $s_{max}=0.18$, which is significantly lower than the capacity in the TT model.  

\begin{figure}%[htbp]
\includegraphics[width=.5\textwidth]{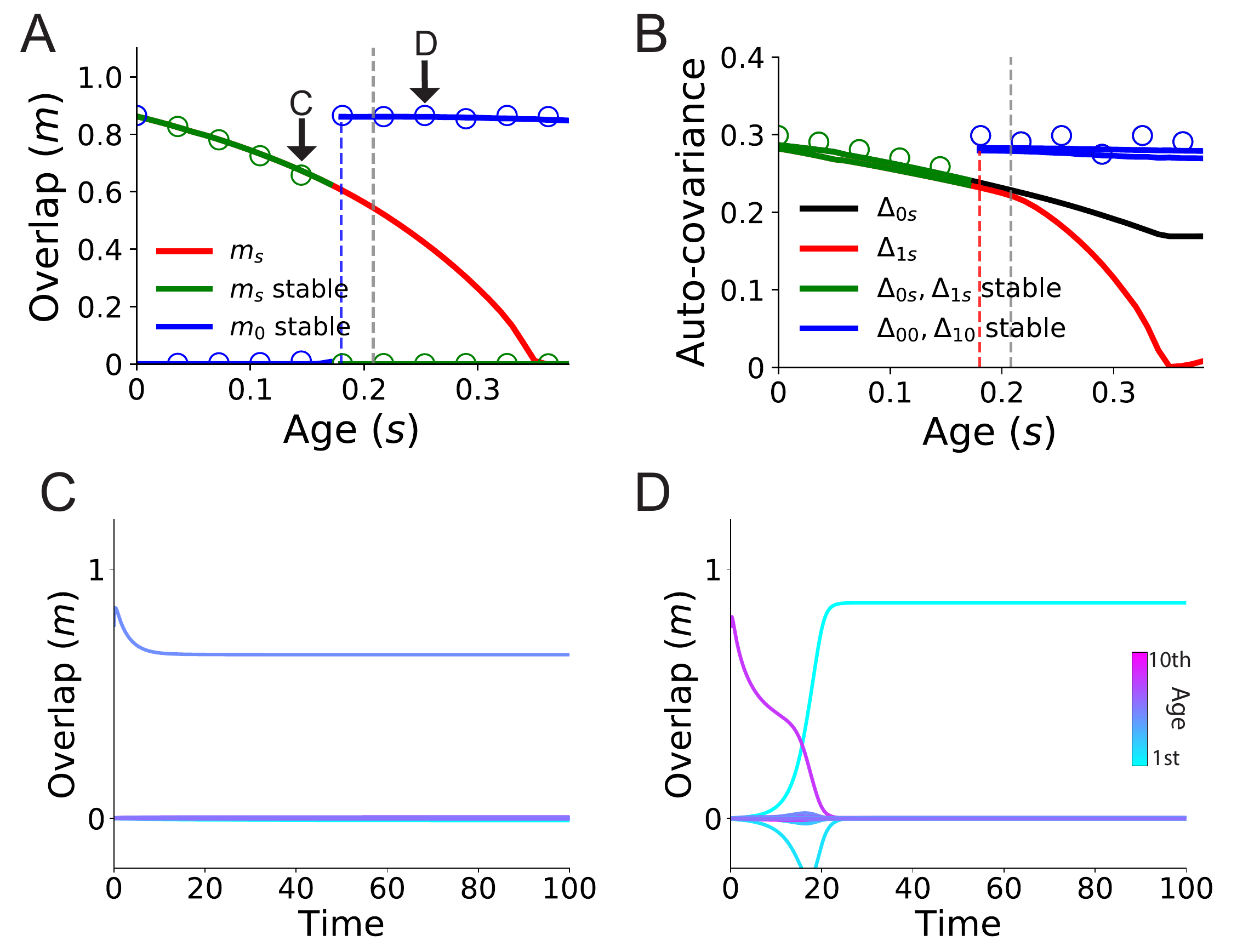}
\caption{(A) Overlap vs memory age $s$ in the network with online learning. Green line/circles: overlap with a memory of age $s$ after the network is initialized close to the corresponding state, calculated from DMFT/
network simulations, respectively. Blue line/circles: overlap with a recent memory that is retrieved instead of the memory of age $s$. Red line: overlap with a memory of age $s$ in the unstable retrieval state (DMFT) In simulations, averages are over time and over 10 realizations. Vertical dashed gray line: transition to chaos. Vertical dashed blue line: Age at which memories become unstable. Red line: overlap with memory of age $s$ (DMFT) Blue line: overlap with most recent memories, $s=0$ (DMFT). (B) Autocovariance parameters vs memory age.  Green line/circles: $\Delta_{0s}=\Delta_{1s}$ computed from DMFT/simulations respectively. Blue line/circles: $\Delta_{00}=\Delta_{10}$ computed from DMFT/simulations respectively. Black/red line: $\Delta_{0s}/\Delta_{1s}$ in unstable retrieval state (DMFT) (C) Dynamics of overlaps for a memory of age $s=5/K =0.18$. The memory is retrieved successfully. (D) Dynamics of overlaps for a memory of age $s=8/K =0.29$. The memory is not retrieved, as the network goes instead in the attractor state corresponding to the most recent memory. In both C and D overlaps are color-coded by age. Network parameters: $N=10^6$, $K= 2 \log(N)$, $\tau=0.64$, and  $A = 4$. }
\label{fig:5}
\end{figure}

\begin{figure}%[htbp]
\includegraphics[width=.5\textwidth]{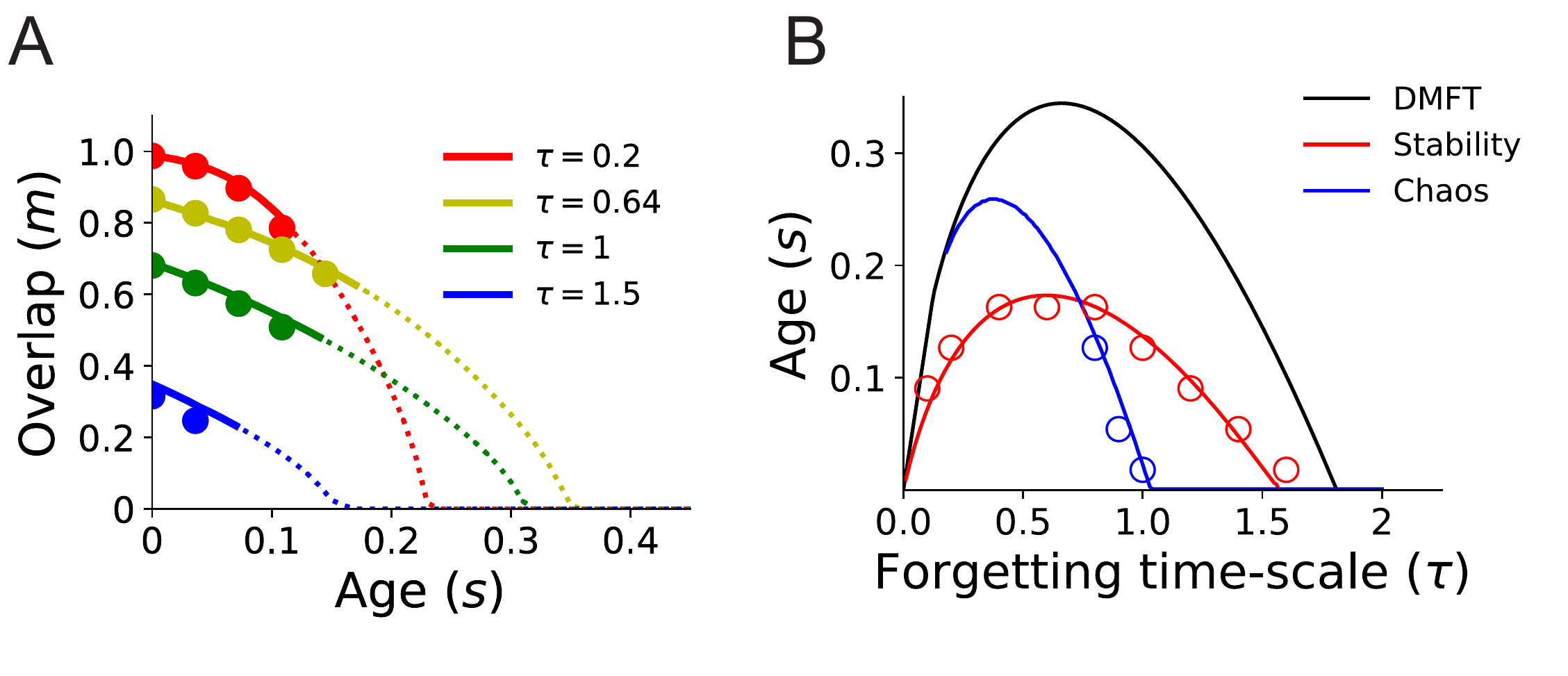}
\caption{(A) Overlap vs memory age ($s$) for different values of the forgetting time scale $\tau$, and $A=4$. Full lines: Overlap computed using DMFT in stable retrieval states. Dashed lines: unstable retrieval states. Circles: Simulations (average over time and over 10 realizations). (B) Bifurcation diagram for the network with forgetting, for $A=4$. Black line: maximal age at which retrieval states exist, calculated using DMFT. Blue line/circles: transition to chaos of retrieval states (DMFT/simulations). Red line/circles: maximal age at which retrieval states are stable (DMFT/simulations). Simulation parameters: $N=10^6$, $K= 2 \log(N)$.}
\label{fig:6}
\end{figure}

Numerical simulations in large very sparsely connected networks show a good agreement with DMFT, as shown in Fig.~\ref{fig:5} in a network with $A=4$ and $\tau=0.64$. Fig.~\ref{fig:5}A,B shows that both the overlap with the retrieved pattern, and the variance $\Delta_{0s}$ decay with age. At the age of about $s=0.18$, retrieval states
become unstable, and the network retrieves one of the most recent memories instead. Fig.~\ref{fig:5}C,D show examples of successful retrieval (Fig.~\ref{fig:5}C), and of an unsuccessful retrieval ending in the retrieval of the most recent memory instead (Fig.~\ref{fig:5}D). Fig.~\ref{fig:6} shows how the overlaps with retrieved memories depend on the forgetting time constant $\tau$. It shows that the overlaps in retrieval states of recent memories decay with $\tau$, since increasing $\tau$ means older patterns provide increasing interference with the retrieval of recent patterns. Finally, Fig.~\ref{fig:6} shows that the location of the two instability lines of retrieval states (in red, the instability towards more recent memories, and in blue, the instability towards chaotic retrieval states) predicted by DMFT are in good agreement with numerical simulations.

\section{Discussion}

\begin{figure}%[htbp]
\includegraphics[width=.48\textwidth]{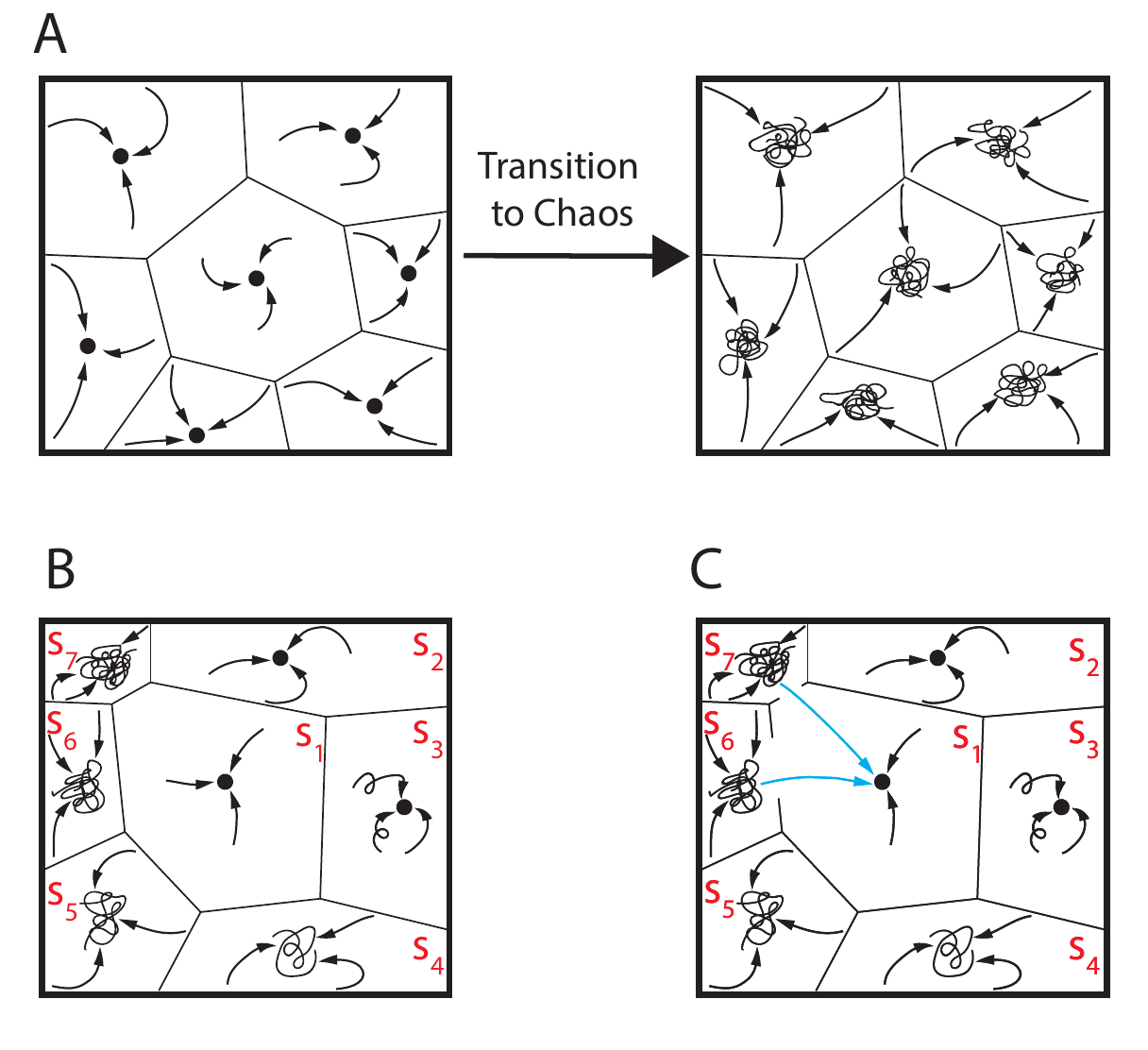}
\caption[Schematics of the attractor networks phase space]{Schematics of the attractor networks phase space. (A) Extensive transition to chaos of fixed-point attractor memory states in an attractor network with no forgetting. When the memory load $\alpha$ surpasses a critical point all the fixed-point attractors transition to chaos at once. For a given value of the memory load $\alpha$, all fixed-point or chaotic memory states are statistically equivalent. (B) In an attractor network that forgets memory states are heterogeneous. Most recent memories are stored as fixed-point attractors with larger basin of attraction while older memories are chaotic with smaller basins of attraction. The chaotic temporal fluctuations become faster for older memories. (C) As predicted by our stability analysis, after a critical memory age, older memory states become unstable and the network jumps to the most recent memory state (see light blue arrows). }
\label{fig:7}
\end{figure}

In the present work, we have characterizez using dynamical mean field theory the dynamics of an attractor network that learns and forgets through an online Hebbian learning rule, and shown that in this network fixed-point and chaotic memory states co-exist.  In such a network, recent memories are strongly imprinted in the connectivity while older memory are exponentially forgotten. In the limit of no forgetting, the network undergoes a global transition to chaos, where all memory states transition to chaos at once when the number of stored patterns surpasses a critical value (see Fig.~\ref{fig:7}A). When forgetting takes place, memory states are heterogeneous and form a continuum of statistically different memory states. In a broad range of forgetting time scales, recent memories are stored in the network as fixed-point attractors, while older memories as chaotic attractors (see Fig.~\ref{fig:7}B). The magnitude of the chaotic fluctuations increase parametrically with the memory age and the basin of attractions shrink with age. Interactions between most recent and older memories destabilize older memory states (see Fig.~\ref{fig:7}C). Our stability analysis explains this effect and accurately predicts the capacity curve, i.e., the curve in parameter space when older memory state destabilize.  We contrast our results with simulation of large networks and find that our theory matches well the network dynamics, provided synaptic connectivity is sparse enough. Overall, our work puts forward a network mechanism that might explain the diversity of neuronal responses in the cortex during memory tasks, and provides an analytical framework for dissecting the network mechanisms of such heterogeneity.

\subsection{Relation with other chaotic networks}

Random networks of rate units are a popular class of tractable models for exploring the neural mechanisms underlying the temporal fluctuations and heterogeneity of the firing rate patterns observed in the brain. Transition to chaos has been extensively studied in randomly connected networks \cite{sompolinsky1988chaos,wainrib2013topological,crisanti2018path,engelken2020lyapunov} and the analysis has been extended to multiple scenarios such as networks with multiple populations \cite{aljadeff2015transition, kadmon2015transition, harish2015asynchronous}, networks of clusters of recurrently connected neurons \cite{stern2014dynamics}, networks driven by external stimuli \cite{rajan2010stimulus, schuecker2018optimal}, and random networks partially structured with low-rank connectivity \cite{mastrogiuseppe2018linking, landau2018coherent,landau2021macroscopic}. The effect of chaos in the computational capabilities of recurrent networks has also been studied extensively \cite{toyoizumi2011beyond,sussillo2009generating,poole2016exponential, schuecker2018optimal,keup2021transient}.  In the vast majority of the above studies the focus has been in the transition to chaos of a single fixed-point, the background (zero average activity) state. A notable exception is the work by \cite{tirozzi1991chaos}. In this work by \citeauthor{tirozzi1991chaos}, they analytically study the highly diluted version of the Hopfield model in a network of continuous `firing rate' units. They derived a DMFT that fully characterizes the networks dynamics. Their DMFT is a particular case of our model for infinite forgetting time scale $\tau \to \infty$ and  number of patterns proportional to $K$, i.e., $p=\alpha K$. The DMFT for the network in \cite{tirozzi1991chaos} is recovered in our equations for $\alpha=\kappa=\tau/2$ and $s=0$. Although in \cite{tirozzi1991chaos} computed analytically the phase diagram of the model, no comparison with network simulations were provided. In Section \ref{sec:tirozzitsodyks} of the present work, we complemented the DMFT in \cite{tirozzi1991chaos}
with a systematic comparison between numerical solutions of the DMFT and simulations of large networks. We show that a good agreement between simulations and theory is only obtained for very sparse networks, in which $c\sim 1/N^a$ and $a>0.8$.  (see Section \ref{sec:discussion:sparsity}). Our work differs from the above studies in that in our network there exist a coexistence between a continuum of statistically distinguishable fixed-point and chaotic memory states (see Fig.~\ref{fig:7} B and C). % The mechanism that gives rise to this diversity is the age-dependence of the mean  and variance input current for each memory state. 

\subsection{Relation with other palimpsest networks}

The problem of catastrophic forgetting was recognized soon after the Hopfield model was proposed, and this issue was addressed using various online learning rules \cite{mezard1986solvable,nadal1986networks,parisi1986memory} that lead to gradual forgetting of old memories, thereby successfully avoiding catastrophic forgetting. This class of models are sometimes referred to as `palimpsest' models. In all these models the price to pay for avoiding catastrophic forgetting is a substantial decrease in storage capacity. This phenomenon is also present in our model, where the capacity is only about $s_{max}=0.18$ for $A=10$, to be compared with a capacity of order 0.9 in the TT model.
%For instance, in the sparsely connected Hopfield model with exponential forgetting, the capacity decreases in a factor of $1/e$ from $\alpha_c=2/\pi = 0.636$ to $s^c=2/(\pi e) =0.234$ for the optimal forgetting time scale \cite{derrida1987learning}. Our model recapitulates these results for the limit $A\to \infty$ when neurons becomes binary as in \cite{derrida1987learning} and attractor states are fixed-points (see Appendix \ref{appendix:optimal}). For chaotic attractors, the optimal capacity slightly increases to $s^c=2/((\pi-2) e)=0.322$ (see Appendix \ref{appendix:optimal}) but it is still  approximately half of what is predicted for the sparse Hopfield model.

Online learning rules has also been widely studied in networks with binary synapses  \cite{tsodyks1990associative,amit1992constraints,amit1994learning,romani2008optimizing,fusi2005cascade,amit2010precise,huang2011capacity,lahiri2013memory, dubreuil2014memory,benna2016computational}. In these models, synapses have two states, one with strong, the other with weak (or zero) efficacy. Transitions between states are induced stochastically by the activity of pre and post-synaptic neurons. Under general conditions, these networks also behave as palimpsests, maintaining a memory of recently shown patterns but gradually forgetting patterns shown in the past. Most studies of such networks have used an ideal observer approach to estimate the maximal age at which a pattern can still be retrieved from synaptic connectivity  \cite{amit1994learning,fusi2005cascade,amit2010precise,lahiri2013memory, benna2016computational}. Storage capacity has also been investigated in attractor networks with synapses, using either binary neurons \cite{amit2010precise,huang2011capacity,dubreuil2014memory}, or firing rate units \cite{romani2008optimizing}. We expect that the results presented here about the diversity of retrieval states should hold in sparsely connected networks with such synapses, but the capacity should depend drastically on the implementation of the Markov process describing the transition between states.

\subsection{Relation to random networks with low-rank structure \label{sec:discussion:lowrank}}

 Recently, low-rank networks have been used as a framework for modeling the dynamics and computations in neuronal circuits in the brain   \cite{mastrogiuseppe2018linking, landau2018coherent, schuessler2020dynamics, beiran2021shaping,landau2021macroscopic}. In these networks, the connectivity is comprised of a low-rank matrix plus a random matrix with i.i.d. Gaussian entries, i.e., $J_{ij}=\sum_{s=1}^p\frac{A}{N}\eta_i^s\xi_j^s +  X_{ij}$ where $X_{ij}\overset{i.i.d.}{\sim}N(0,g^2/N)$ and $p\sim\mathcal{O}(1)$.  Attractor networks and low-rank networks are very similar from a mean field perspective. In the attractor network analyzed here, for a given memory state $s$, in the large $N$ and $K$ limit, the result of our DMFT is equivalent to the following low-rank network $\frac{A}{N}\eta_i^s\eta_j^se^{-s/\tau} + \frac{A\tau}{2\sqrt{N}} X_{ij}$, where $X_{ij}\overset{i.i.d.}{\sim}N(0,1)$. Therefore, the mean input current due to the low-rank component in low-rank networks is equivalent to the mean input current of the retrieved memory in attractor networks, while the random component  is equivalent to the quenched noise that arises from the stored memories (see \cite{beiran2021shaping}). 
 %A key difference is that attractor networks  capacities scale linearly with the network size while in low-rank networks it is usually much smaller. 
 %The equivalence between low-rank networks with multiple populations and the Hopfield model in the zero memory load limit has been recently shown \cite{beiran2021shaping}. 
 Because these networks are similar in the mean field limit, we hypothesize that by introducing heterogeneity in the strength of low-rank components, the co-existence between chaotic and fixed-point attractors displayed in our network can also be obtained in low-rank networks.

\subsection{Network sparsity \label{sec:discussion:sparsity}}
%\cite{martin1973statistical,de1978field,janssen1976lagrangean, sompolinsky1982relaxational, helias2020statistical}
The DMFT derived here, as well as the ones in \cite{tsodyks1988associative, tirozzi1991chaos}, can be obtained using a  path integral formalism, expanding the generating functional in powers of $c=K/N\ll 1$, and then neglecting the  $\mathcal{O}(K^2/N^2)$ corrections \cite{kree1987continuous}. However, the conditions on $c$ and $N$ for which these higher order corrections can be neglected remained unclear. Our numerical simulations show that our DMFT accurately predicts the network dynamics in very sparse networks with $c\sim\log(N)/N$ (Fig.~\ref{fig:3}). For denser connectivity, the agreement between capacity predicted by theory and simulations degrades progressively (see Fig.~\ref{fig:3} A and B). On the other hand, the transition to chaos is well predicted by theory up to connection probabilities $c\sim 1/N^{0.4}$, at which point the chaotic region disappears altogether. These results are consistent with results on sparse networks of binary neurons where the disruptive effect of correlations between neurons in the capacity can be avoided when $K\ll \log(N)$ \cite{derrida1987exactly}.

\subsection{Attractor networks with learning rules inferred from data}

We have recently studied an attractor network similar to the one presented here, but with both transfer function and synaptic plasticity rules inferred from \emph{in vivo} data \cite{pereira2018attractor}.
We showed that this network display qualitatively similar dynamics as  the TT model. For small learning gain $A$ and memory loads $\alpha$ memory states are fixed-point attractors, while for large $A$ and $\alpha$ memory states are chaotic (see Fig.~4 and Fig.~6 respectively in \cite{pereira2018attractor}). A detailed study of this more realistic network with an online learning rule is outside the scope of this paper.
%Since learning rules inferred from data produce extremely sparse patterns \cite{lim2015inferring, pereira2018attractor} we expect that simulating these networks in the highly sparse connectivity limit $c\sim \log(N)/N$ (where our DMFT is exact) will be very difficult to achieve, needing extremely large networks for obtaining a good match with network simulations. 
Building and analyzing more realistic attractor network models with learning rules inferred from \emph{in vivo} data should help to clarify the network mechanisms underlying the observed changes in neural activity across days during learning novel images \cite{huang2018neural, garrett2020experience}.  % It would be interesting 

%We need to Discuss what we say here.

%Discuss how this model relates with Pereira \& Brunel, 18. We have to address why we are not doing the DMFT for this model and the palimpsest version in this paper without giving away the mixed-state results and telling the whole mixed-states story. It would be great to be able to say something as many old attractor network papers say: `this will studied elsewhere'. But I think in the current times we cannot longer do that. 
 %I have investigated what happens in the highly-sparse limit log(N)/N in a network with learning rules inferred from data. For the network sizes I am able to simulate (~1M neurons) in the highly sparse limit these networks are not able to retrieve memories. In the beginning, I was convinced I had a bug in my code, but then I tried with denser sparsities as 1/sqrt(N) and these networks retrieved memories successfully. I didn't understand the reason for this. But I think I have an explanation. Since K is very small and the learning rule produces very sparse patterns, if we think of storing one binary pattern, then the high_i *high_j entries are very few, if we multiply high_i *high_j  by the sparse connectivity then is very unlikely that for a finite network the entries high_i *high_j (where c_ij=1) are actually observed, although it is possible in the large N limit. Then for finite networks, the highly sparse patterns and highly sparse connectivity limit it is very difficult to achieve via simulations.

\subsection{Experimental predictions}

To investigate the neural mechanisms underlying the maintenance and manipulation of information stored in memory, neuroscientists have designed a broad class of tasks named delay response tasks. In these tasks, subjects must keep in memory information about a previously presented stimulus during a delay period, to be able to produce a behavioral response that depends on this information. 
Neuronal recordings in delay response tasks show that a (typically small proportion) of neurons display elevated tonic activity during the delay period \cite{fuster1971neuron, miyashita1988neuronal, funahashi1989mnemonic, goldman1995cellular}. The observed persistent elevated delay activity is consistent with attractor network models in which fixed-point attractors correspond the memory states \cite{amit1997model, wang1999synaptic}. However, in many of  these recordings, neurons exhibit firing patterns with high degree of temporal variation. These observations challenge attractor networks as viable models of memory storage \cite{lundqvist2018working}. Interestingly, population analysis of neural activity during delay response tasks shows that in spite of the variability observed at a single neuron level there is a stable population encoding of the memoranda during the delay periods \cite{murray2017stable}. A possible scenario compatible with the above observations \cite{pereira2018attractor, aljadeff2021synapse} is that the single neuron temporal variability and stable population encoding can be explained by attractor networks in the chaotic regime as described here for the TT model in Section \ref{sec:tirozzitsodyks} (see Fig.~\ref{fig:1}; green region and Fig.~\ref{fig:7}A).   In an attractor network with forgetting, memory states are heterogeneous and as a memory ages, its activity ranges from fixed-point to chaotic. Additionally, for chaotic retrieval states, the temporal fluctuations of their activity gets faster with age (see Fig.~\ref{fig:6}B). Our theory thus predicts that for delay response tasks with aging memoranda: 1) retrieval states corresponding to newer memories will present a large proportion of neurons with persistent delay activity consistent with fixed-point attractor dynamics; 2) temporal fluctuations will become more pronounced, and faster, as memories become older. These predictions can be tested by recording neural activity during delay response tasks performed across the learning and forgetting process of multiple items. 
%To this moment we are not aware of such experiments. We would be excited to test the experimental predictions of our theory in future research.

\section{Acknowledgments}
UP thanks to the Swartz Fundation for its support and to the Champaign Public Library for providing the physical space where an early version of this work was developed. 
NB was supported by R01MH115555, R01NS112917 and ONR N00014-17-1-3004.

%%%%%%%%%%%%%%%%%%%%%%%%%%%%%%%%%%%%%%%%%%
%%%%%%%%%Methods
%%%%%%%%%%%%%%%%%%%%%%%%%%%%%%%%%%%%%%%%%%

\appendix

\section{\label{appendix:relaxational}Relaxational dynamics for $\Delta_{0s}$ and $\Delta_{1s}$}

For computing the order parameters $m_0$, $m_s$, $\Delta_{0s}$ and $\Delta_{1s}$ in panels Fig.~\ref{fig:5} A (green and blue lines) and B (red and yellow lines), we replace the full integro-differential  dynamics  of  the  variance  $\Delta_{0s}$ and  the  two-time autocorrelation  function (see section 2.2 in the supplementary information of \cite{gillett2020characteristics} for the corresponding equations in a closely related model)    by  a relaxational dynamics for  $\Delta_{0s}$ and $\Delta_{1s}$ given by

 \begin{widetext}
  \begin{equation}
  \begin{split}
\label{appendinx:rel:1}
\frac{{\rm d}\Delta_{1s}}{{\rm d}t} = -\Delta_{1s} + \frac{\tau}{4}  \int Dz\left\{ \left[\int Dx \phi\left(A\left [ \sqrt{\Delta_{0s}-|\Delta_{1s}|} x+\sqrt{|\Delta_{1s}|} z+m_0+e^{-\frac{s}{\tau}}m_{s}\right]\right) \right]^2+\right. \\
\left. \left[\int Dx \phi\left(A\left [ \sqrt{\Delta_{0s}-|\Delta_{1s}|} x+\sqrt{|\Delta_{1s}|} z+e^{-\frac{s}{\tau}}m_{s}-m_0\right]\right) \right]^2\right\}
\end{split}
\end{equation}
\end{widetext}

and

\begin{widetext}
 \begin{equation}
\label{appendinx:rel:2}
\begin{split}
&\frac{{\rm d}\Delta_{0s}}{{\rm d}t}  = -\Delta_{0s} +\sqrt{\Delta_{1s}^2+\psi_1(m_{0},m_{s}, \Delta_{0s}, \Delta_{1s})-\psi_2(m_{0},m_{s}, \Delta_{0s}, \Delta_{1s})},
\end{split}
\end{equation}
\end{widetext}

with 

\begin{widetext}
 \begin{equation}
\label{appendinx:rel:3}
\begin{split}
\psi_1(m_{0},m_{s}, \Delta_{0s}, \Delta_{1s}) = \frac{\tau }{2A^2}  \int  D x \left\{  \Phi^2\left(A\left[\sqrt{\Delta_{0s}} x+m_0+e^{-\frac{s}{\tau}}m_{s}\right]\right)+\Phi^2\left(A\left[\sqrt{\Delta_{0s}} x+e^{-\frac{s}{\tau}}m_{s}-m_0\right]\right)\right\}
\end{split}
\end{equation}
\end{widetext}

\begin{widetext}
 \begin{equation}
\label{appendinx:rel:4}
\begin{split}
&\psi_2(m_{0}, m_{s}, \Delta_{0s}, \Delta_{1s}) = \frac{\tau}{ 2A^2}\int   Dz\left\{ \left[\int Dx \Phi\left(A\left [ \sqrt{\Delta_{0s}-|\Delta_{1s}|} x+\sqrt{|\Delta_1|} z +m_0+e^{-\frac{s}{\tau}}m_{s}\right]\right)\right]^2+\right.\\
&\left.\left[\int Dx \Phi\left(A\left [ \sqrt{\Delta_{0s}-|\Delta_{1s}|} x+\sqrt{|\Delta_{1s}|} z+e^{-\frac{s}{\tau}}m_{s}-m_0\right]\right)\right]^2\right\}.
\end{split}
\end{equation}
\end{widetext}

Notice that the dynamics for the overlaps is given by Eqs.~(\ref{amendeddmft:5}, \ref{amendeddmft:6}). Therefore, the full dynamics for the order parameters $m_{0}, m_{s}, \Delta_{0s}, \Delta_{1s}$ is ruled by a system of integro-differential equations given by Eqs.~(\ref{appendinx:rel:1}, \ref{appendinx:rel:2}, \ref{amendeddmft:5}, \ref{amendeddmft:6}). We initialize the dynamics using several initial conditions of the the form 

\begin{eqnarray}
m_0 &=& \delta_{m_0}\\
m_{s} &=& 1\\
\Delta_{0s}& = & d_0\\
\Delta_{1s} &= &d_1,
\end{eqnarray}

where $d_0$ and $d_1$ are fixed and $\delta_{m_0}$ is small and variable. The steady state solutions are shown in Fig.~\ref{fig:5} A and B. This method makes an accurate  prediction for the memory age $s^*$ above which networks jump form an older to the most recent memory state (see Fig.~\ref{fig:5} A and B). %The implementation of this method is publicly available (see next section).

\bigskip

\section{Network simulations and numerical solutions to the DMFT}

The network simulations and numerical solutions to the DMFT were generated using custom Python scripts.  This code is available at GitHub repository  \href{https://github.com/ulisespereira/chaos-forgetting-palimpsest}{https://github.com/ulisespereira/chaos-forgetting-palimpsest}.

 %----------------------------------------------
%--------References------------------------
%-----------------------------------------------

\bibliography{references.bib}

\begin{thebibliography}{80}%
\makeatletter
\providecommand \@ifxundefined [1]{%
 \@ifx{#1\undefined}
}%
\providecommand \@ifnum [1]{%
 \ifnum #1\expandafter \@firstoftwo
 \else \expandafter \@secondoftwo
 \fi
}%
\providecommand \@ifx [1]{%
 \ifx #1\expandafter \@firstoftwo
 \else \expandafter \@secondoftwo
 \fi
}%
\providecommand \natexlab [1]{#1}%
\providecommand \enquote  [1]{``#1''}%
\providecommand \bibnamefont  [1]{#1}%
\providecommand \bibfnamefont [1]{#1}%
\providecommand \citenamefont [1]{#1}%
\providecommand \href@noop [0]{\@secondoftwo}%
\providecommand \href [0]{\begingroup \@sanitize@url \@href}%
\providecommand \@href[1]{\@@startlink{#1}\@@href}%
\providecommand \@@href[1]{\endgroup#1\@@endlink}%
\providecommand \@sanitize@url [0]{\catcode `\\12\catcode `\$12\catcode
  `\&12\catcode `\#12\catcode `\^12\catcode `\_12\catcode `\%12\relax}%
\providecommand \@@startlink[1]{}%
\providecommand \@@endlink[0]{}%
\providecommand \url  [0]{\begingroup\@sanitize@url \@url }%
\providecommand \@url [1]{\endgroup\@href {#1}{\urlprefix }}%
\providecommand \urlprefix  [0]{URL }%
\providecommand \Eprint [0]{\href }%
\providecommand \doibase [0]{https://doi.org/}%
\providecommand \selectlanguage [0]{\@gobble}%
\providecommand \bibinfo  [0]{\@secondoftwo}%
\providecommand \bibfield  [0]{\@secondoftwo}%
\providecommand \translation [1]{[#1]}%
\providecommand \BibitemOpen [0]{}%
\providecommand \bibitemStop [0]{}%
\providecommand \bibitemNoStop [0]{.\EOS\space}%
\providecommand \EOS [0]{\spacefactor3000\relax}%
\providecommand \BibitemShut  [1]{\csname bibitem#1\endcsname}%
\let\auto@bib@innerbib\@empty
%</preamble>
\bibitem [{\citenamefont {Hopfield}(1982)}]{hopfield1982neural}%
  \BibitemOpen
  \bibfield  {author} {\bibinfo {author} {\bibfnamefont {J.~J.}\ \bibnamefont
  {Hopfield}},\ }\bibfield  {title} {\bibinfo {title} {Neural networks and
  physical systems with emergent collective computational abilities},\
  }\href@noop {} {\bibfield  {journal} {\bibinfo  {journal} {Proceedings of the
  national academy of sciences}\ }\textbf {\bibinfo {volume} {79}},\ \bibinfo
  {pages} {2554} (\bibinfo {year} {1982})}\BibitemShut {NoStop}%
\bibitem [{\citenamefont {Amit}\ \emph {et~al.}(1985)\citenamefont {Amit},
  \citenamefont {Gutfreund},\ and\ \citenamefont {Sompolinsky}}]{amit1985spin}%
  \BibitemOpen
  \bibfield  {author} {\bibinfo {author} {\bibfnamefont {D.~J.}\ \bibnamefont
  {Amit}}, \bibinfo {author} {\bibfnamefont {H.}~\bibnamefont {Gutfreund}},\
  and\ \bibinfo {author} {\bibfnamefont {H.}~\bibnamefont {Sompolinsky}},\
  }\bibfield  {title} {\bibinfo {title} {Spin-glass models of neural
  networks},\ }\href@noop {} {\bibfield  {journal} {\bibinfo  {journal}
  {Physical Review A}\ }\textbf {\bibinfo {volume} {32}},\ \bibinfo {pages}
  {1007} (\bibinfo {year} {1985})}\BibitemShut {NoStop}%
\bibitem [{\citenamefont {Amit}(1992)}]{amit1992modeling}%
  \BibitemOpen
  \bibfield  {author} {\bibinfo {author} {\bibfnamefont {D.~J.}\ \bibnamefont
  {Amit}},\ }\href@noop {} {\emph {\bibinfo {title} {Modeling brain function:
  The world of attractor neural networks}}}\ (\bibinfo  {publisher} {Cambridge
  University Press},\ \bibinfo {year} {1992})\BibitemShut {NoStop}%
\bibitem [{\citenamefont {Brunel}(2005)}]{brunel2005network}%
  \BibitemOpen
  \bibfield  {author} {\bibinfo {author} {\bibfnamefont {N.}~\bibnamefont
  {Brunel}},\ }\bibfield  {title} {\bibinfo {title} {Network models of
  memory},\ }in\ \href@noop {} {\emph {\bibinfo {booktitle} {Methods and Models
  in Neurophysics, Volume Session LXXX: Lecture Notes of the Les Houches Summer
  School 2003}}},\ \bibinfo {editor} {edited by\ \bibinfo {editor}
  {\bibfnamefont {C.}~\bibnamefont {Chow}}, \bibinfo {editor} {\bibfnamefont
  {B.}~\bibnamefont {Gutkin}}, \bibinfo {editor} {\bibfnamefont
  {D.}~\bibnamefont {Hansel}}, \bibinfo {editor} {\bibfnamefont
  {C.}~\bibnamefont {Meunier}},\ and\ \bibinfo {editor} {\bibfnamefont
  {J.}~\bibnamefont {Dalibard}}}\ (\bibinfo  {publisher} {Elsevier},\ \bibinfo
  {year} {2005})\ pp.\ \bibinfo {pages} {407--476}\BibitemShut {NoStop}%
\bibitem [{\citenamefont {Tirozzi}\ and\ \citenamefont
  {Tsodyks}(1991)}]{tirozzi1991chaos}%
  \BibitemOpen
  \bibfield  {author} {\bibinfo {author} {\bibfnamefont {B.}~\bibnamefont
  {Tirozzi}}\ and\ \bibinfo {author} {\bibfnamefont {M.}~\bibnamefont
  {Tsodyks}},\ }\bibfield  {title} {\bibinfo {title} {Chaos in highly diluted
  neural networks},\ }\href@noop {} {\bibfield  {journal} {\bibinfo  {journal}
  {EPL (Europhysics Letters)}\ }\textbf {\bibinfo {volume} {14}},\ \bibinfo
  {pages} {727} (\bibinfo {year} {1991})}\BibitemShut {NoStop}%
\bibitem [{\citenamefont {Fuster}\ \emph {et~al.}(1971)\citenamefont {Fuster},
  \citenamefont {Alexander} \emph {et~al.}}]{fuster1971neuron}%
  \BibitemOpen
  \bibfield  {author} {\bibinfo {author} {\bibfnamefont {J.~M.}\ \bibnamefont
  {Fuster}}, \bibinfo {author} {\bibfnamefont {G.~E.}\ \bibnamefont
  {Alexander}}, \emph {et~al.},\ }\bibfield  {title} {\bibinfo {title} {Neuron
  activity related to short-term memory},\ }\href@noop {} {\bibfield  {journal}
  {\bibinfo  {journal} {Science}\ }\textbf {\bibinfo {volume} {173}},\ \bibinfo
  {pages} {652} (\bibinfo {year} {1971})}\BibitemShut {NoStop}%
\bibitem [{\citenamefont {Miyashita}(1988)}]{miyashita1988neuronal}%
  \BibitemOpen
  \bibfield  {author} {\bibinfo {author} {\bibfnamefont {Y.}~\bibnamefont
  {Miyashita}},\ }\bibfield  {title} {\bibinfo {title} {Neuronal correlate of
  visual associative long-term memory in the primate temporal cortex},\
  }\href@noop {} {\bibfield  {journal} {\bibinfo  {journal} {Nature}\ }\textbf
  {\bibinfo {volume} {335}},\ \bibinfo {pages} {817} (\bibinfo {year}
  {1988})}\BibitemShut {NoStop}%
\bibitem [{\citenamefont {Funahashi}\ \emph {et~al.}(1989)\citenamefont
  {Funahashi}, \citenamefont {Bruce},\ and\ \citenamefont
  {Goldman-Rakic}}]{funahashi1989mnemonic}%
  \BibitemOpen
  \bibfield  {author} {\bibinfo {author} {\bibfnamefont {S.}~\bibnamefont
  {Funahashi}}, \bibinfo {author} {\bibfnamefont {C.~J.}\ \bibnamefont
  {Bruce}},\ and\ \bibinfo {author} {\bibfnamefont {P.~S.}\ \bibnamefont
  {Goldman-Rakic}},\ }\bibfield  {title} {\bibinfo {title} {Mnemonic coding of
  visual space in the monkey's dorsolateral prefrontal cortex},\ }\href@noop {}
  {\bibfield  {journal} {\bibinfo  {journal} {Journal of neurophysiology}\
  }\textbf {\bibinfo {volume} {61}},\ \bibinfo {pages} {331} (\bibinfo {year}
  {1989})}\BibitemShut {NoStop}%
\bibitem [{\citenamefont {Goldman-Rakic}(1995)}]{goldman1995cellular}%
  \BibitemOpen
  \bibfield  {author} {\bibinfo {author} {\bibfnamefont {P.~S.}\ \bibnamefont
  {Goldman-Rakic}},\ }\bibfield  {title} {\bibinfo {title} {Cellular basis of
  working memory},\ }\href@noop {} {\bibfield  {journal} {\bibinfo  {journal}
  {Neuron}\ }\textbf {\bibinfo {volume} {14}},\ \bibinfo {pages} {477}
  (\bibinfo {year} {1995})}\BibitemShut {NoStop}%
\bibitem [{\citenamefont {Liu}\ \emph {et~al.}(2014)\citenamefont {Liu},
  \citenamefont {Gu}, \citenamefont {Zhu}, \citenamefont {Zhang}, \citenamefont
  {Han}, \citenamefont {Yan}, \citenamefont {Cheng}, \citenamefont {Hao},
  \citenamefont {Fan}, \citenamefont {Hou} \emph {et~al.}}]{liu2014medial}%
  \BibitemOpen
  \bibfield  {author} {\bibinfo {author} {\bibfnamefont {D.}~\bibnamefont
  {Liu}}, \bibinfo {author} {\bibfnamefont {X.}~\bibnamefont {Gu}}, \bibinfo
  {author} {\bibfnamefont {J.}~\bibnamefont {Zhu}}, \bibinfo {author}
  {\bibfnamefont {X.}~\bibnamefont {Zhang}}, \bibinfo {author} {\bibfnamefont
  {Z.}~\bibnamefont {Han}}, \bibinfo {author} {\bibfnamefont {W.}~\bibnamefont
  {Yan}}, \bibinfo {author} {\bibfnamefont {Q.}~\bibnamefont {Cheng}}, \bibinfo
  {author} {\bibfnamefont {J.}~\bibnamefont {Hao}}, \bibinfo {author}
  {\bibfnamefont {H.}~\bibnamefont {Fan}}, \bibinfo {author} {\bibfnamefont
  {R.}~\bibnamefont {Hou}}, \emph {et~al.},\ }\bibfield  {title} {\bibinfo
  {title} {Medial prefrontal activity during delay period contributes to
  learning of a working memory task},\ }\href@noop {} {\bibfield  {journal}
  {\bibinfo  {journal} {Science}\ }\textbf {\bibinfo {volume} {346}},\ \bibinfo
  {pages} {458} (\bibinfo {year} {2014})}\BibitemShut {NoStop}%
\bibitem [{\citenamefont {Guo}\ \emph {et~al.}(2014)\citenamefont {Guo},
  \citenamefont {Li}, \citenamefont {Huber}, \citenamefont {Ophir},
  \citenamefont {Gutnisky}, \citenamefont {Ting}, \citenamefont {Feng},\ and\
  \citenamefont {Svoboda}}]{guo2014flow}%
  \BibitemOpen
  \bibfield  {author} {\bibinfo {author} {\bibfnamefont {Z.~V.}\ \bibnamefont
  {Guo}}, \bibinfo {author} {\bibfnamefont {N.}~\bibnamefont {Li}}, \bibinfo
  {author} {\bibfnamefont {D.}~\bibnamefont {Huber}}, \bibinfo {author}
  {\bibfnamefont {E.}~\bibnamefont {Ophir}}, \bibinfo {author} {\bibfnamefont
  {D.}~\bibnamefont {Gutnisky}}, \bibinfo {author} {\bibfnamefont {J.~T.}\
  \bibnamefont {Ting}}, \bibinfo {author} {\bibfnamefont {G.}~\bibnamefont
  {Feng}},\ and\ \bibinfo {author} {\bibfnamefont {K.}~\bibnamefont
  {Svoboda}},\ }\bibfield  {title} {\bibinfo {title} {Flow of cortical activity
  underlying a tactile decision in mice},\ }\href@noop {} {\bibfield  {journal}
  {\bibinfo  {journal} {Neuron}\ }\textbf {\bibinfo {volume} {81}},\ \bibinfo
  {pages} {179} (\bibinfo {year} {2014})}\BibitemShut {NoStop}%
\bibitem [{\citenamefont {Inagaki}\ \emph {et~al.}(2017)\citenamefont
  {Inagaki}, \citenamefont {Fontolan}, \citenamefont {Romani},\ and\
  \citenamefont {Svoboda}}]{inagaki2017discrete}%
  \BibitemOpen
  \bibfield  {author} {\bibinfo {author} {\bibfnamefont {H.~K.}\ \bibnamefont
  {Inagaki}}, \bibinfo {author} {\bibfnamefont {L.}~\bibnamefont {Fontolan}},
  \bibinfo {author} {\bibfnamefont {S.}~\bibnamefont {Romani}},\ and\ \bibinfo
  {author} {\bibfnamefont {K.}~\bibnamefont {Svoboda}},\ }\bibfield  {title}
  {\bibinfo {title} {Discrete attractor dynamics underlying selective
  persistent activity in frontal cortex},\ }\href@noop {} {\bibfield  {journal}
  {\bibinfo  {journal} {Biorxiv}\ ,\ \bibinfo {pages} {203448}} (\bibinfo
  {year} {2017})}\BibitemShut {NoStop}%
\bibitem [{\citenamefont {Compte}\ \emph {et~al.}(2003)\citenamefont {Compte},
  \citenamefont {Constantinidis}, \citenamefont {Tegn\'er}, \citenamefont
  {Raghavachari}, \citenamefont {Chafee}, \citenamefont {Goldman-Rakic},\ and\
  \citenamefont {Wang}}]{compte2003temporally}%
  \BibitemOpen
  \bibfield  {author} {\bibinfo {author} {\bibfnamefont {A.}~\bibnamefont
  {Compte}}, \bibinfo {author} {\bibfnamefont {C.}~\bibnamefont
  {Constantinidis}}, \bibinfo {author} {\bibfnamefont {J.}~\bibnamefont
  {Tegn\'er}}, \bibinfo {author} {\bibfnamefont {S.}~\bibnamefont
  {Raghavachari}}, \bibinfo {author} {\bibfnamefont {M.}~\bibnamefont
  {Chafee}}, \bibinfo {author} {\bibfnamefont {P.~S.}\ \bibnamefont
  {Goldman-Rakic}},\ and\ \bibinfo {author} {\bibfnamefont {X.-J.}\
  \bibnamefont {Wang}},\ }\bibfield  {title} {\bibinfo {title} {Temporally
  irregular mnemonic persistent activity in prefrontal neurons of monkeys
  during a delayed response task},\ }\href@noop {} {\bibfield  {journal}
  {\bibinfo  {journal} {J.~Neurophysiol.}\ }\textbf {\bibinfo {volume} {90}},\
  \bibinfo {pages} {3441} (\bibinfo {year} {2003})}\BibitemShut {NoStop}%
\bibitem [{\citenamefont {Shafi}\ \emph {et~al.}(2007)\citenamefont {Shafi},
  \citenamefont {Zhou}, \citenamefont {Quintana}, \citenamefont {Chow},
  \citenamefont {Fuster},\ and\ \citenamefont {Bodner}}]{shafi2007variability}%
  \BibitemOpen
  \bibfield  {author} {\bibinfo {author} {\bibfnamefont {M.}~\bibnamefont
  {Shafi}}, \bibinfo {author} {\bibfnamefont {Y.}~\bibnamefont {Zhou}},
  \bibinfo {author} {\bibfnamefont {J.}~\bibnamefont {Quintana}}, \bibinfo
  {author} {\bibfnamefont {C.}~\bibnamefont {Chow}}, \bibinfo {author}
  {\bibfnamefont {J.}~\bibnamefont {Fuster}},\ and\ \bibinfo {author}
  {\bibfnamefont {M.}~\bibnamefont {Bodner}},\ }\bibfield  {title} {\bibinfo
  {title} {Variability in neuronal activity in primate cortex during working
  memory tasks},\ }\href@noop {} {\bibfield  {journal} {\bibinfo  {journal}
  {Neuroscience}\ }\textbf {\bibinfo {volume} {146}},\ \bibinfo {pages} {1082}
  (\bibinfo {year} {2007})}\BibitemShut {NoStop}%
\bibitem [{\citenamefont {Barak}\ \emph {et~al.}(2010)\citenamefont {Barak},
  \citenamefont {Tsodyks},\ and\ \citenamefont {Romo}}]{barak2010neuronal}%
  \BibitemOpen
  \bibfield  {author} {\bibinfo {author} {\bibfnamefont {O.}~\bibnamefont
  {Barak}}, \bibinfo {author} {\bibfnamefont {M.}~\bibnamefont {Tsodyks}},\
  and\ \bibinfo {author} {\bibfnamefont {R.}~\bibnamefont {Romo}},\ }\bibfield
  {title} {\bibinfo {title} {{{N}euronal population coding of parametric
  working memory}},\ }\href@noop {} {\bibfield  {journal} {\bibinfo  {journal}
  {J. Neurosci.}\ }\textbf {\bibinfo {volume} {30}},\ \bibinfo {pages} {9424}
  (\bibinfo {year} {2010})}\BibitemShut {NoStop}%
\bibitem [{\citenamefont {Barak}\ and\ \citenamefont
  {Tsodyks}(2014)}]{barak2014working}%
  \BibitemOpen
  \bibfield  {author} {\bibinfo {author} {\bibfnamefont {O.}~\bibnamefont
  {Barak}}\ and\ \bibinfo {author} {\bibfnamefont {M.}~\bibnamefont
  {Tsodyks}},\ }\bibfield  {title} {\bibinfo {title} {{{W}orking models of
  working memory}},\ }\href@noop {} {\bibfield  {journal} {\bibinfo  {journal}
  {Curr. Opin. Neurobiol.}\ }\textbf {\bibinfo {volume} {25}},\ \bibinfo
  {pages} {20} (\bibinfo {year} {2014})}\BibitemShut {NoStop}%
\bibitem [{\citenamefont {Kobak}\ \emph {et~al.}(2016)\citenamefont {Kobak},
  \citenamefont {Brendel}, \citenamefont {Constantinidis}, \citenamefont
  {Feierstein}, \citenamefont {Kepecs}, \citenamefont {Mainen}, \citenamefont
  {Qi}, \citenamefont {Romo}, \citenamefont {Uchida},\ and\ \citenamefont
  {Machens}}]{kobak2016demixed}%
  \BibitemOpen
  \bibfield  {author} {\bibinfo {author} {\bibfnamefont {D.}~\bibnamefont
  {Kobak}}, \bibinfo {author} {\bibfnamefont {W.}~\bibnamefont {Brendel}},
  \bibinfo {author} {\bibfnamefont {C.}~\bibnamefont {Constantinidis}},
  \bibinfo {author} {\bibfnamefont {C.~E.}\ \bibnamefont {Feierstein}},
  \bibinfo {author} {\bibfnamefont {A.}~\bibnamefont {Kepecs}}, \bibinfo
  {author} {\bibfnamefont {Z.~F.}\ \bibnamefont {Mainen}}, \bibinfo {author}
  {\bibfnamefont {X.~L.}\ \bibnamefont {Qi}}, \bibinfo {author} {\bibfnamefont
  {R.}~\bibnamefont {Romo}}, \bibinfo {author} {\bibfnamefont {N.}~\bibnamefont
  {Uchida}},\ and\ \bibinfo {author} {\bibfnamefont {C.~K.}\ \bibnamefont
  {Machens}},\ }\bibfield  {title} {\bibinfo {title} {{{D}emixed principal
  component analysis of neural population data}},\ }\href@noop {} {\bibfield
  {journal} {\bibinfo  {journal} {Elife}\ }\textbf {\bibinfo {volume} {5}}
  (\bibinfo {year} {2016})}\BibitemShut {NoStop}%
\bibitem [{\citenamefont {Murray}\ \emph {et~al.}(2017)\citenamefont {Murray},
  \citenamefont {Bernacchia}, \citenamefont {Roy}, \citenamefont
  {Constantinidis}, \citenamefont {Romo},\ and\ \citenamefont
  {Wang}}]{murray2017stable}%
  \BibitemOpen
  \bibfield  {author} {\bibinfo {author} {\bibfnamefont {J.~D.}\ \bibnamefont
  {Murray}}, \bibinfo {author} {\bibfnamefont {A.}~\bibnamefont {Bernacchia}},
  \bibinfo {author} {\bibfnamefont {N.~A.}\ \bibnamefont {Roy}}, \bibinfo
  {author} {\bibfnamefont {C.}~\bibnamefont {Constantinidis}}, \bibinfo
  {author} {\bibfnamefont {R.}~\bibnamefont {Romo}},\ and\ \bibinfo {author}
  {\bibfnamefont {X.-J.}\ \bibnamefont {Wang}},\ }\bibfield  {title} {\bibinfo
  {title} {Stable population coding for working memory coexists with
  heterogeneous neural dynamics in prefrontal cortex},\ }\href@noop {}
  {\bibfield  {journal} {\bibinfo  {journal} {Proceedings of the National
  Academy of Sciences}\ }\textbf {\bibinfo {volume} {114}},\ \bibinfo {pages}
  {394} (\bibinfo {year} {2017})}\BibitemShut {NoStop}%
\bibitem [{\citenamefont {Barbieri}\ and\ \citenamefont
  {Brunel}(2007)}]{barbieri2007irregular}%
  \BibitemOpen
  \bibfield  {author} {\bibinfo {author} {\bibfnamefont {F.}~\bibnamefont
  {Barbieri}}\ and\ \bibinfo {author} {\bibfnamefont {N.}~\bibnamefont
  {Brunel}},\ }\bibfield  {title} {\bibinfo {title} {Irregular persistent
  activity induced by synaptic excitatory feedback},\ }\href@noop {} {\bibfield
   {journal} {\bibinfo  {journal} {Frontiers in Computational Neuroscience}\
  }\textbf {\bibinfo {volume} {1}},\ \bibinfo {pages} {5} (\bibinfo {year}
  {2007})}\BibitemShut {NoStop}%
\bibitem [{\citenamefont {Mongillo}\ \emph {et~al.}(2008)\citenamefont
  {Mongillo}, \citenamefont {Barak},\ and\ \citenamefont
  {Tsodyks}}]{mongillo2008synaptic}%
  \BibitemOpen
  \bibfield  {author} {\bibinfo {author} {\bibfnamefont {G.}~\bibnamefont
  {Mongillo}}, \bibinfo {author} {\bibfnamefont {O.}~\bibnamefont {Barak}},\
  and\ \bibinfo {author} {\bibfnamefont {M.}~\bibnamefont {Tsodyks}},\
  }\bibfield  {title} {\bibinfo {title} {Synaptic theory of working memory},\
  }\href@noop {} {\bibfield  {journal} {\bibinfo  {journal} {Science}\ }\textbf
  {\bibinfo {volume} {319}},\ \bibinfo {pages} {1543} (\bibinfo {year}
  {2008})}\BibitemShut {NoStop}%
\bibitem [{\citenamefont {Lundqvist}\ \emph {et~al.}(2010)\citenamefont
  {Lundqvist}, \citenamefont {Compte},\ and\ \citenamefont
  {Lansner}}]{lundqvist2010bistable}%
  \BibitemOpen
  \bibfield  {author} {\bibinfo {author} {\bibfnamefont {M.}~\bibnamefont
  {Lundqvist}}, \bibinfo {author} {\bibfnamefont {A.}~\bibnamefont {Compte}},\
  and\ \bibinfo {author} {\bibfnamefont {A.}~\bibnamefont {Lansner}},\
  }\bibfield  {title} {\bibinfo {title} {{{B}istable, irregular firing and
  population oscillations in a modular attractor memory network}},\ }\href@noop
  {} {\bibfield  {journal} {\bibinfo  {journal} {PLoS Comput. Biol.}\ }\textbf
  {\bibinfo {volume} {6}},\ \bibinfo {pages} {e1000803} (\bibinfo {year}
  {2010})}\BibitemShut {NoStop}%
\bibitem [{\citenamefont {Druckmann}\ and\ \citenamefont
  {Chklovskii}(2012)}]{druckmann2012neuronal}%
  \BibitemOpen
  \bibfield  {author} {\bibinfo {author} {\bibfnamefont {S.}~\bibnamefont
  {Druckmann}}\ and\ \bibinfo {author} {\bibfnamefont {D.~B.}\ \bibnamefont
  {Chklovskii}},\ }\bibfield  {title} {\bibinfo {title} {Neuronal circuits
  underlying persistent representations despite time varying activity},\
  }\href@noop {} {\bibfield  {journal} {\bibinfo  {journal} {Current Biology}\
  }\textbf {\bibinfo {volume} {22}},\ \bibinfo {pages} {2095} (\bibinfo {year}
  {2012})}\BibitemShut {NoStop}%
\bibitem [{\citenamefont {Pereira}\ and\ \citenamefont
  {Brunel}(2018)}]{pereira2018attractor}%
  \BibitemOpen
  \bibfield  {author} {\bibinfo {author} {\bibfnamefont {U.}~\bibnamefont
  {Pereira}}\ and\ \bibinfo {author} {\bibfnamefont {N.}~\bibnamefont
  {Brunel}},\ }\bibfield  {title} {\bibinfo {title} {Attractor dynamics in
  networks with learning rules inferred from in vivo data},\ }\href@noop {}
  {\bibfield  {journal} {\bibinfo  {journal} {Neuron}\ }\textbf {\bibinfo
  {volume} {99}},\ \bibinfo {pages} {227} (\bibinfo {year} {2018})}\BibitemShut
  {NoStop}%
\bibitem [{\citenamefont {Tsodyks}\ and\ \citenamefont
  {Feigel'Man}(1988)}]{tsodyks1988enhanced}%
  \BibitemOpen
  \bibfield  {author} {\bibinfo {author} {\bibfnamefont {M.}~\bibnamefont
  {Tsodyks}}\ and\ \bibinfo {author} {\bibfnamefont {M.}~\bibnamefont
  {Feigel'Man}},\ }\bibfield  {title} {\bibinfo {title} {The enhanced storage
  capacity in neural networks with low activity level},\ }\href@noop {}
  {\bibfield  {journal} {\bibinfo  {journal} {EPL (Europhysics Letters)}\
  }\textbf {\bibinfo {volume} {6}},\ \bibinfo {pages} {101} (\bibinfo {year}
  {1988})}\BibitemShut {NoStop}%
\bibitem [{\citenamefont {Sompolinsky}\ \emph {et~al.}(1988)\citenamefont
  {Sompolinsky}, \citenamefont {Crisanti},\ and\ \citenamefont
  {Sommers}}]{sompolinsky1988chaos}%
  \BibitemOpen
  \bibfield  {author} {\bibinfo {author} {\bibfnamefont {H.}~\bibnamefont
  {Sompolinsky}}, \bibinfo {author} {\bibfnamefont {A.}~\bibnamefont
  {Crisanti}},\ and\ \bibinfo {author} {\bibfnamefont {H.-J.}\ \bibnamefont
  {Sommers}},\ }\bibfield  {title} {\bibinfo {title} {Chaos in random neural
  networks},\ }\href@noop {} {\bibfield  {journal} {\bibinfo  {journal}
  {Physical Review Letters}\ }\textbf {\bibinfo {volume} {61}},\ \bibinfo
  {pages} {259} (\bibinfo {year} {1988})}\BibitemShut {NoStop}%
\bibitem [{\citenamefont {Crisanti}\ and\ \citenamefont
  {Sompolinsky}(2018)}]{crisanti2018path}%
  \BibitemOpen
  \bibfield  {author} {\bibinfo {author} {\bibfnamefont {A.}~\bibnamefont
  {Crisanti}}\ and\ \bibinfo {author} {\bibfnamefont {H.}~\bibnamefont
  {Sompolinsky}},\ }\bibfield  {title} {\bibinfo {title} {Path integral
  approach to random neural networks},\ }\href@noop {} {\bibfield  {journal}
  {\bibinfo  {journal} {Physical Review E}\ }\textbf {\bibinfo {volume} {98}},\
  \bibinfo {pages} {062120} (\bibinfo {year} {2018})}\BibitemShut {NoStop}%
\bibitem [{\citenamefont {Nadal}\ \emph {et~al.}(1986)\citenamefont {Nadal},
  \citenamefont {Toulouse}, \citenamefont {Changeux},\ and\ \citenamefont
  {Dehaene}}]{nadal1986networks}%
  \BibitemOpen
  \bibfield  {author} {\bibinfo {author} {\bibfnamefont {J.}~\bibnamefont
  {Nadal}}, \bibinfo {author} {\bibfnamefont {G.}~\bibnamefont {Toulouse}},
  \bibinfo {author} {\bibfnamefont {J.}~\bibnamefont {Changeux}},\ and\
  \bibinfo {author} {\bibfnamefont {S.}~\bibnamefont {Dehaene}},\ }\bibfield
  {title} {\bibinfo {title} {Networks of formal neurons and memory
  palimpsests},\ }\href@noop {} {\bibfield  {journal} {\bibinfo  {journal} {EPL
  (Europhysics Letters)}\ }\textbf {\bibinfo {volume} {1}},\ \bibinfo {pages}
  {535} (\bibinfo {year} {1986})}\BibitemShut {NoStop}%
\bibitem [{\citenamefont {Parisi}(1986)}]{parisi1986memory}%
  \BibitemOpen
  \bibfield  {author} {\bibinfo {author} {\bibfnamefont {G.}~\bibnamefont
  {Parisi}},\ }\bibfield  {title} {\bibinfo {title} {A memory which forgets},\
  }\href@noop {} {\bibfield  {journal} {\bibinfo  {journal} {Journal of Physics
  A: Mathematical and General}\ }\textbf {\bibinfo {volume} {19}},\ \bibinfo
  {pages} {L617} (\bibinfo {year} {1986})}\BibitemShut {NoStop}%
\bibitem [{\citenamefont {M{\'e}zard}\ \emph {et~al.}(1986)\citenamefont
  {M{\'e}zard}, \citenamefont {Nadal},\ and\ \citenamefont
  {Toulouse}}]{mezard1986solvable}%
  \BibitemOpen
  \bibfield  {author} {\bibinfo {author} {\bibfnamefont {M.}~\bibnamefont
  {M{\'e}zard}}, \bibinfo {author} {\bibfnamefont {J.}~\bibnamefont {Nadal}},\
  and\ \bibinfo {author} {\bibfnamefont {G.}~\bibnamefont {Toulouse}},\
  }\bibfield  {title} {\bibinfo {title} {Solvable models of working memories},\
  }\href@noop {} {\bibfield  {journal} {\bibinfo  {journal} {Journal de
  physique}\ }\textbf {\bibinfo {volume} {47}},\ \bibinfo {pages} {1457}
  (\bibinfo {year} {1986})}\BibitemShut {NoStop}%
\bibitem [{\citenamefont {Tsodyks}(1990)}]{tsodyks1990associative}%
  \BibitemOpen
  \bibfield  {author} {\bibinfo {author} {\bibfnamefont {M.}~\bibnamefont
  {Tsodyks}},\ }\bibfield  {title} {\bibinfo {title} {Associative memory in
  neural networks with binary synapses},\ }\href@noop {} {\bibfield  {journal}
  {\bibinfo  {journal} {Modern Physics Letters B}\ }\textbf {\bibinfo {volume}
  {4}},\ \bibinfo {pages} {713} (\bibinfo {year} {1990})}\BibitemShut {NoStop}%
\bibitem [{\citenamefont {Amit}\ and\ \citenamefont
  {Fusi}(1994)}]{amit1994learning}%
  \BibitemOpen
  \bibfield  {author} {\bibinfo {author} {\bibfnamefont {D.~J.}\ \bibnamefont
  {Amit}}\ and\ \bibinfo {author} {\bibfnamefont {S.}~\bibnamefont {Fusi}},\
  }\bibfield  {title} {\bibinfo {title} {Learning in neural networks with
  material synapses},\ }\href@noop {} {\bibfield  {journal} {\bibinfo
  {journal} {Neural Computation}\ }\textbf {\bibinfo {volume} {6}},\ \bibinfo
  {pages} {957} (\bibinfo {year} {1994})}\BibitemShut {NoStop}%
\bibitem [{\citenamefont {Fusi}\ \emph {et~al.}(2005)\citenamefont {Fusi},
  \citenamefont {Drew},\ and\ \citenamefont {Abbott}}]{fusi2005cascade}%
  \BibitemOpen
  \bibfield  {author} {\bibinfo {author} {\bibfnamefont {S.}~\bibnamefont
  {Fusi}}, \bibinfo {author} {\bibfnamefont {P.~J.}\ \bibnamefont {Drew}},\
  and\ \bibinfo {author} {\bibfnamefont {L.~F.}\ \bibnamefont {Abbott}},\
  }\bibfield  {title} {\bibinfo {title} {Cascade models of synaptically stored
  memories},\ }\href@noop {} {\bibfield  {journal} {\bibinfo  {journal}
  {Neuron}\ }\textbf {\bibinfo {volume} {45}},\ \bibinfo {pages} {599}
  (\bibinfo {year} {2005})}\BibitemShut {NoStop}%
\bibitem [{\citenamefont {Fusi}\ and\ \citenamefont
  {Abbott}(2007)}]{fusi2007limits}%
  \BibitemOpen
  \bibfield  {author} {\bibinfo {author} {\bibfnamefont {S.}~\bibnamefont
  {Fusi}}\ and\ \bibinfo {author} {\bibfnamefont {L.}~\bibnamefont {Abbott}},\
  }\bibfield  {title} {\bibinfo {title} {Limits on the memory storage capacity
  of bounded synapses},\ }\href@noop {} {\bibfield  {journal} {\bibinfo
  {journal} {Nature neuroscience}\ }\textbf {\bibinfo {volume} {10}},\ \bibinfo
  {pages} {485} (\bibinfo {year} {2007})}\BibitemShut {NoStop}%
\bibitem [{\citenamefont {Romani}\ \emph {et~al.}(2008)\citenamefont {Romani},
  \citenamefont {Amit},\ and\ \citenamefont {Amit}}]{romani2008optimizing}%
  \BibitemOpen
  \bibfield  {author} {\bibinfo {author} {\bibfnamefont {S.}~\bibnamefont
  {Romani}}, \bibinfo {author} {\bibfnamefont {D.~J.}\ \bibnamefont {Amit}},\
  and\ \bibinfo {author} {\bibfnamefont {Y.}~\bibnamefont {Amit}},\ }\bibfield
  {title} {\bibinfo {title} {Optimizing one-shot learning with binary
  synapses},\ }\href@noop {} {\bibfield  {journal} {\bibinfo  {journal} {Neural
  computation}\ }\textbf {\bibinfo {volume} {20}},\ \bibinfo {pages} {1928}
  (\bibinfo {year} {2008})}\BibitemShut {NoStop}%
\bibitem [{\citenamefont {Dubreuil}\ \emph {et~al.}(2014)\citenamefont
  {Dubreuil}, \citenamefont {Amit},\ and\ \citenamefont
  {Brunel}}]{dubreuil2014memory}%
  \BibitemOpen
  \bibfield  {author} {\bibinfo {author} {\bibfnamefont {A.~M.}\ \bibnamefont
  {Dubreuil}}, \bibinfo {author} {\bibfnamefont {Y.}~\bibnamefont {Amit}},\
  and\ \bibinfo {author} {\bibfnamefont {N.}~\bibnamefont {Brunel}},\
  }\bibfield  {title} {\bibinfo {title} {Memory capacity of networks with
  stochastic binary synapses},\ }\href@noop {} {\bibfield  {journal} {\bibinfo
  {journal} {PLoS computational biology}\ }\textbf {\bibinfo {volume} {10}},\
  \bibinfo {pages} {e1003727} (\bibinfo {year} {2014})}\BibitemShut {NoStop}%
\bibitem [{\citenamefont {Huang}\ and\ \citenamefont
  {Amit}(2011)}]{huang2011capacity}%
  \BibitemOpen
  \bibfield  {author} {\bibinfo {author} {\bibfnamefont {Y.}~\bibnamefont
  {Huang}}\ and\ \bibinfo {author} {\bibfnamefont {Y.}~\bibnamefont {Amit}},\
  }\bibfield  {title} {\bibinfo {title} {Capacity analysis in multi-state
  synaptic models: a retrieval probability perspective},\ }\href@noop {}
  {\bibfield  {journal} {\bibinfo  {journal} {Journal of computational
  neuroscience}\ }\textbf {\bibinfo {volume} {30}},\ \bibinfo {pages} {699}
  (\bibinfo {year} {2011})}\BibitemShut {NoStop}%
\bibitem [{\citenamefont {Lahiri}\ and\ \citenamefont
  {Ganguli}(2013)}]{lahiri2013memory}%
  \BibitemOpen
  \bibfield  {author} {\bibinfo {author} {\bibfnamefont {S.}~\bibnamefont
  {Lahiri}}\ and\ \bibinfo {author} {\bibfnamefont {S.}~\bibnamefont
  {Ganguli}},\ }\bibfield  {title} {\bibinfo {title} {A memory frontier for
  complex synapses},\ }in\ \href@noop {} {\emph {\bibinfo {booktitle} {Advances
  in neural information processing systems}}}\ (\bibinfo {year} {2013})\ pp.\
  \bibinfo {pages} {1034--1042}\BibitemShut {NoStop}%
\bibitem [{\citenamefont {Benna}\ and\ \citenamefont
  {Fusi}(2016)}]{benna2016computational}%
  \BibitemOpen
  \bibfield  {author} {\bibinfo {author} {\bibfnamefont {M.~K.}\ \bibnamefont
  {Benna}}\ and\ \bibinfo {author} {\bibfnamefont {S.}~\bibnamefont {Fusi}},\
  }\bibfield  {title} {\bibinfo {title} {Computational principles of synaptic
  memory consolidation},\ }\href@noop {} {\bibfield  {journal} {\bibinfo
  {journal} {Nature neuroscience}\ }\textbf {\bibinfo {volume} {19}},\ \bibinfo
  {pages} {1697} (\bibinfo {year} {2016})}\BibitemShut {NoStop}%
\bibitem [{\citenamefont {Ostojic}\ and\ \citenamefont
  {Brunel}(2011)}]{ostojic2011spiking}%
  \BibitemOpen
  \bibfield  {author} {\bibinfo {author} {\bibfnamefont {S.}~\bibnamefont
  {Ostojic}}\ and\ \bibinfo {author} {\bibfnamefont {N.}~\bibnamefont
  {Brunel}},\ }\bibfield  {title} {\bibinfo {title} {From spiking neuron models
  to linear-nonlinear models},\ }\href@noop {} {\bibfield  {journal} {\bibinfo
  {journal} {PLoS Comput Biol}\ }\textbf {\bibinfo {volume} {7}},\ \bibinfo
  {pages} {e1001056} (\bibinfo {year} {2011})}\BibitemShut {NoStop}%
\bibitem [{\citenamefont {Mason}\ \emph {et~al.}(1991)\citenamefont {Mason},
  \citenamefont {Nicoll},\ and\ \citenamefont {Stratford}}]{mason1991synaptic}%
  \BibitemOpen
  \bibfield  {author} {\bibinfo {author} {\bibfnamefont {A.}~\bibnamefont
  {Mason}}, \bibinfo {author} {\bibfnamefont {A.}~\bibnamefont {Nicoll}},\ and\
  \bibinfo {author} {\bibfnamefont {K.}~\bibnamefont {Stratford}},\ }\bibfield
  {title} {\bibinfo {title} {Synaptic transmission between individual pyramidal
  neurons of the rat visual cortex in vitro},\ }\href@noop {} {\bibfield
  {journal} {\bibinfo  {journal} {Journal of Neuroscience}\ }\textbf {\bibinfo
  {volume} {11}},\ \bibinfo {pages} {72} (\bibinfo {year} {1991})}\BibitemShut
  {NoStop}%
\bibitem [{\citenamefont {Markram}\ \emph {et~al.}(1997)\citenamefont
  {Markram}, \citenamefont {L{\"u}bke}, \citenamefont {Frotscher},
  \citenamefont {Roth},\ and\ \citenamefont {Sakmann}}]{markram1997physiology}%
  \BibitemOpen
  \bibfield  {author} {\bibinfo {author} {\bibfnamefont {H.}~\bibnamefont
  {Markram}}, \bibinfo {author} {\bibfnamefont {J.}~\bibnamefont {L{\"u}bke}},
  \bibinfo {author} {\bibfnamefont {M.}~\bibnamefont {Frotscher}}, \bibinfo
  {author} {\bibfnamefont {A.}~\bibnamefont {Roth}},\ and\ \bibinfo {author}
  {\bibfnamefont {B.}~\bibnamefont {Sakmann}},\ }\bibfield  {title} {\bibinfo
  {title} {Physiology and anatomy of synaptic connections between thick tufted
  pyramidal neurones in the developing rat neocortex.},\ }\href@noop {}
  {\bibfield  {journal} {\bibinfo  {journal} {The Journal of physiology}\
  }\textbf {\bibinfo {volume} {500}},\ \bibinfo {pages} {409} (\bibinfo {year}
  {1997})}\BibitemShut {NoStop}%
\bibitem [{\citenamefont {Holmgren}\ \emph {et~al.}(2003)\citenamefont
  {Holmgren}, \citenamefont {Harkany}, \citenamefont {Svennenfors},\ and\
  \citenamefont {Zilberter}}]{holmgren2003pyramidal}%
  \BibitemOpen
  \bibfield  {author} {\bibinfo {author} {\bibfnamefont {C.}~\bibnamefont
  {Holmgren}}, \bibinfo {author} {\bibfnamefont {T.}~\bibnamefont {Harkany}},
  \bibinfo {author} {\bibfnamefont {B.}~\bibnamefont {Svennenfors}},\ and\
  \bibinfo {author} {\bibfnamefont {Y.}~\bibnamefont {Zilberter}},\ }\bibfield
  {title} {\bibinfo {title} {Pyramidal cell communication within local networks
  in layer 2/3 of rat neocortex},\ }\href@noop {} {\bibfield  {journal}
  {\bibinfo  {journal} {The Journal of physiology}\ }\textbf {\bibinfo {volume}
  {551}},\ \bibinfo {pages} {139} (\bibinfo {year} {2003})}\BibitemShut
  {NoStop}%
\bibitem [{\citenamefont {Thomson}\ and\ \citenamefont
  {Lamy}(2007)}]{thomson2007functional}%
  \BibitemOpen
  \bibfield  {author} {\bibinfo {author} {\bibfnamefont {A.~M.}\ \bibnamefont
  {Thomson}}\ and\ \bibinfo {author} {\bibfnamefont {C.}~\bibnamefont {Lamy}},\
  }\bibfield  {title} {\bibinfo {title} {Functional maps of neocortical local
  circuitry},\ }\href@noop {} {\bibfield  {journal} {\bibinfo  {journal}
  {Frontiers in neuroscience}\ }\textbf {\bibinfo {volume} {1}},\ \bibinfo
  {pages} {2} (\bibinfo {year} {2007})}\BibitemShut {NoStop}%
\bibitem [{\citenamefont {Lefort}\ \emph {et~al.}(2009)\citenamefont {Lefort},
  \citenamefont {Tomm}, \citenamefont {Sarria},\ and\ \citenamefont
  {Petersen}}]{lefort2009excitatory}%
  \BibitemOpen
  \bibfield  {author} {\bibinfo {author} {\bibfnamefont {S.}~\bibnamefont
  {Lefort}}, \bibinfo {author} {\bibfnamefont {C.}~\bibnamefont {Tomm}},
  \bibinfo {author} {\bibfnamefont {J.-C.~F.}\ \bibnamefont {Sarria}},\ and\
  \bibinfo {author} {\bibfnamefont {C.~C.}\ \bibnamefont {Petersen}},\
  }\bibfield  {title} {\bibinfo {title} {The excitatory neuronal network of the
  c2 barrel column in mouse primary somatosensory cortex},\ }\href@noop {}
  {\bibfield  {journal} {\bibinfo  {journal} {Neuron}\ }\textbf {\bibinfo
  {volume} {61}},\ \bibinfo {pages} {301} (\bibinfo {year} {2009})}\BibitemShut
  {NoStop}%
\bibitem [{\citenamefont {Guzman}\ \emph {et~al.}(2016)\citenamefont {Guzman},
  \citenamefont {Schl{\"o}gl}, \citenamefont {Frotscher},\ and\ \citenamefont
  {Jonas}}]{guzman2016synaptic}%
  \BibitemOpen
  \bibfield  {author} {\bibinfo {author} {\bibfnamefont {S.~J.}\ \bibnamefont
  {Guzman}}, \bibinfo {author} {\bibfnamefont {A.}~\bibnamefont {Schl{\"o}gl}},
  \bibinfo {author} {\bibfnamefont {M.}~\bibnamefont {Frotscher}},\ and\
  \bibinfo {author} {\bibfnamefont {P.}~\bibnamefont {Jonas}},\ }\bibfield
  {title} {\bibinfo {title} {Synaptic mechanisms of pattern completion in the
  hippocampal ca3 network},\ }\href@noop {} {\bibfield  {journal} {\bibinfo
  {journal} {Science}\ }\textbf {\bibinfo {volume} {353}},\ \bibinfo {pages}
  {1117} (\bibinfo {year} {2016})}\BibitemShut {NoStop}%
\bibitem [{\citenamefont {Sejnowski}(1977)}]{sejnowski1977storing}%
  \BibitemOpen
  \bibfield  {author} {\bibinfo {author} {\bibfnamefont {T.~J.}\ \bibnamefont
  {Sejnowski}},\ }\bibfield  {title} {\bibinfo {title} {Storing covariance with
  nonlinearly interacting neurons},\ }\href@noop {} {\bibfield  {journal}
  {\bibinfo  {journal} {Journal of mathematical biology}\ }\textbf {\bibinfo
  {volume} {4}},\ \bibinfo {pages} {303} (\bibinfo {year} {1977})}\BibitemShut
  {NoStop}%
\bibitem [{\citenamefont {Kree}\ and\ \citenamefont
  {Zippelius}(1987)}]{kree1987continuous}%
  \BibitemOpen
  \bibfield  {author} {\bibinfo {author} {\bibfnamefont {R.}~\bibnamefont
  {Kree}}\ and\ \bibinfo {author} {\bibfnamefont {A.}~\bibnamefont
  {Zippelius}},\ }\bibfield  {title} {\bibinfo {title} {Continuous-time
  dynamics of asymmetrically diluted neural networks},\ }\href@noop {}
  {\bibfield  {journal} {\bibinfo  {journal} {Physical Review A}\ }\textbf
  {\bibinfo {volume} {36}},\ \bibinfo {pages} {4421} (\bibinfo {year}
  {1987})}\BibitemShut {NoStop}%
\bibitem [{\citenamefont {Toyoizumi}\ \emph {et~al.}(2014)\citenamefont
  {Toyoizumi}, \citenamefont {Kaneko}, \citenamefont {Stryker},\ and\
  \citenamefont {Miller}}]{toyoizumi2014modeling}%
  \BibitemOpen
  \bibfield  {author} {\bibinfo {author} {\bibfnamefont {T.}~\bibnamefont
  {Toyoizumi}}, \bibinfo {author} {\bibfnamefont {M.}~\bibnamefont {Kaneko}},
  \bibinfo {author} {\bibfnamefont {M.~P.}\ \bibnamefont {Stryker}},\ and\
  \bibinfo {author} {\bibfnamefont {K.~D.}\ \bibnamefont {Miller}},\ }\bibfield
   {title} {\bibinfo {title} {Modeling the dynamic interaction of hebbian and
  homeostatic plasticity},\ }\href@noop {} {\bibfield  {journal} {\bibinfo
  {journal} {Neuron}\ }\textbf {\bibinfo {volume} {84}},\ \bibinfo {pages}
  {497} (\bibinfo {year} {2014})}\BibitemShut {NoStop}%
\bibitem [{\citenamefont {Vogels}\ \emph {et~al.}(2011)\citenamefont {Vogels},
  \citenamefont {Sprekeler}, \citenamefont {Zenke}, \citenamefont {Clopath},\
  and\ \citenamefont {Gerstner}}]{vogels2011inhibitory}%
  \BibitemOpen
  \bibfield  {author} {\bibinfo {author} {\bibfnamefont {T.}~\bibnamefont
  {Vogels}}, \bibinfo {author} {\bibfnamefont {H.}~\bibnamefont {Sprekeler}},
  \bibinfo {author} {\bibfnamefont {F.}~\bibnamefont {Zenke}}, \bibinfo
  {author} {\bibfnamefont {C.}~\bibnamefont {Clopath}},\ and\ \bibinfo {author}
  {\bibfnamefont {W.}~\bibnamefont {Gerstner}},\ }\bibfield  {title} {\bibinfo
  {title} {Inhibitory plasticity balances excitation and inhibition in sensory
  pathways and memory networks},\ }\href@noop {} {\bibfield  {journal}
  {\bibinfo  {journal} {Science}\ }\textbf {\bibinfo {volume} {334}},\ \bibinfo
  {pages} {1569} (\bibinfo {year} {2011})}\BibitemShut {NoStop}%
\bibitem [{\citenamefont {Kadmon}\ and\ \citenamefont
  {Sompolinsky}(2015)}]{kadmon2015transition}%
  \BibitemOpen
  \bibfield  {author} {\bibinfo {author} {\bibfnamefont {J.}~\bibnamefont
  {Kadmon}}\ and\ \bibinfo {author} {\bibfnamefont {H.}~\bibnamefont
  {Sompolinsky}},\ }\bibfield  {title} {\bibinfo {title} {Transition to chaos
  in random neuronal networks},\ }\href@noop {} {\bibfield  {journal} {\bibinfo
   {journal} {Physical Review X}\ }\textbf {\bibinfo {volume} {5}},\ \bibinfo
  {pages} {041030} (\bibinfo {year} {2015})}\BibitemShut {NoStop}%
\bibitem [{\citenamefont {Sch{\"u}cker}\ \emph {et~al.}(2016)\citenamefont
  {Sch{\"u}cker}, \citenamefont {Goedeke}, \citenamefont {Dahmen},\ and\
  \citenamefont {Helias}}]{schucker2016functional}%
  \BibitemOpen
  \bibfield  {author} {\bibinfo {author} {\bibfnamefont {J.}~\bibnamefont
  {Sch{\"u}cker}}, \bibinfo {author} {\bibfnamefont {S.}~\bibnamefont
  {Goedeke}}, \bibinfo {author} {\bibfnamefont {D.}~\bibnamefont {Dahmen}},\
  and\ \bibinfo {author} {\bibfnamefont {M.}~\bibnamefont {Helias}},\
  }\bibfield  {title} {\bibinfo {title} {Functional methods for disordered
  neural networks},\ }\href@noop {} {\bibfield  {journal} {\bibinfo  {journal}
  {arXiv preprint arXiv:1605.06758}\ } (\bibinfo {year} {2016})}\BibitemShut
  {NoStop}%
\bibitem [{\citenamefont {Gillett}\ \emph {et~al.}(2020)\citenamefont
  {Gillett}, \citenamefont {Pereira},\ and\ \citenamefont
  {Brunel}}]{gillett2020characteristics}%
  \BibitemOpen
  \bibfield  {author} {\bibinfo {author} {\bibfnamefont {M.}~\bibnamefont
  {Gillett}}, \bibinfo {author} {\bibfnamefont {U.}~\bibnamefont {Pereira}},\
  and\ \bibinfo {author} {\bibfnamefont {N.}~\bibnamefont {Brunel}},\
  }\bibfield  {title} {\bibinfo {title} {Characteristics of sequential activity
  in networks with temporally asymmetric hebbian learning},\ }\href@noop {}
  {\bibfield  {journal} {\bibinfo  {journal} {Proceedings of the National
  Academy of Sciences}\ }\textbf {\bibinfo {volume} {117}},\ \bibinfo {pages}
  {29948} (\bibinfo {year} {2020})}\BibitemShut {NoStop}%
\bibitem [{\citenamefont {K{\"u}hn}\ \emph {et~al.}(1991)\citenamefont
  {K{\"u}hn}, \citenamefont {B{\"o}s},\ and\ \citenamefont {van
  Hemmen}}]{kuhn1991statistical}%
  \BibitemOpen
  \bibfield  {author} {\bibinfo {author} {\bibfnamefont {R.}~\bibnamefont
  {K{\"u}hn}}, \bibinfo {author} {\bibfnamefont {S.}~\bibnamefont {B{\"o}s}},\
  and\ \bibinfo {author} {\bibfnamefont {J.~L.}\ \bibnamefont {van Hemmen}},\
  }\bibfield  {title} {\bibinfo {title} {Statistical mechanics for networks of
  graded-response neurons},\ }\href@noop {} {\bibfield  {journal} {\bibinfo
  {journal} {Physical Review A}\ }\textbf {\bibinfo {volume} {43}},\ \bibinfo
  {pages} {2084} (\bibinfo {year} {1991})}\BibitemShut {NoStop}%
\bibitem [{\citenamefont {Hopfield}(1984)}]{hopfield1984neurons}%
  \BibitemOpen
  \bibfield  {author} {\bibinfo {author} {\bibfnamefont {J.~J.}\ \bibnamefont
  {Hopfield}},\ }\bibfield  {title} {\bibinfo {title} {Neurons with graded
  response have collective computational properties like those of two-state
  neurons},\ }\href@noop {} {\bibfield  {journal} {\bibinfo  {journal}
  {Proc.~Natl.~Acad.~Sci.~U.S.A.}\ }\textbf {\bibinfo {volume} {81}},\ \bibinfo
  {pages} {3088} (\bibinfo {year} {1984})}\BibitemShut {NoStop}%
\bibitem [{\citenamefont {Wainrib}\ and\ \citenamefont
  {Touboul}(2013)}]{wainrib2013topological}%
  \BibitemOpen
  \bibfield  {author} {\bibinfo {author} {\bibfnamefont {G.}~\bibnamefont
  {Wainrib}}\ and\ \bibinfo {author} {\bibfnamefont {J.}~\bibnamefont
  {Touboul}},\ }\bibfield  {title} {\bibinfo {title} {Topological and dynamical
  complexity of random neural networks},\ }\href@noop {} {\bibfield  {journal}
  {\bibinfo  {journal} {Physical review letters}\ }\textbf {\bibinfo {volume}
  {110}},\ \bibinfo {pages} {118101} (\bibinfo {year} {2013})}\BibitemShut
  {NoStop}%
\bibitem [{\citenamefont {Engelken}\ \emph {et~al.}(2020)\citenamefont
  {Engelken}, \citenamefont {Wolf},\ and\ \citenamefont
  {Abbott}}]{engelken2020lyapunov}%
  \BibitemOpen
  \bibfield  {author} {\bibinfo {author} {\bibfnamefont {R.}~\bibnamefont
  {Engelken}}, \bibinfo {author} {\bibfnamefont {F.}~\bibnamefont {Wolf}},\
  and\ \bibinfo {author} {\bibfnamefont {L.}~\bibnamefont {Abbott}},\
  }\bibfield  {title} {\bibinfo {title} {Lyapunov spectra of chaotic recurrent
  neural networks},\ }\href@noop {} {\bibfield  {journal} {\bibinfo  {journal}
  {arXiv preprint arXiv:2006.02427}\ } (\bibinfo {year} {2020})}\BibitemShut
  {NoStop}%
\bibitem [{\citenamefont {Aljadeff}\ \emph {et~al.}(2015)\citenamefont
  {Aljadeff}, \citenamefont {Stern},\ and\ \citenamefont
  {Sharpee}}]{aljadeff2015transition}%
  \BibitemOpen
  \bibfield  {author} {\bibinfo {author} {\bibfnamefont {J.}~\bibnamefont
  {Aljadeff}}, \bibinfo {author} {\bibfnamefont {M.}~\bibnamefont {Stern}},\
  and\ \bibinfo {author} {\bibfnamefont {T.}~\bibnamefont {Sharpee}},\
  }\bibfield  {title} {\bibinfo {title} {Transition to chaos in random networks
  with cell-type-specific connectivity},\ }\href@noop {} {\bibfield  {journal}
  {\bibinfo  {journal} {Physical review letters}\ }\textbf {\bibinfo {volume}
  {114}},\ \bibinfo {pages} {088101} (\bibinfo {year} {2015})}\BibitemShut
  {NoStop}%
\bibitem [{\citenamefont {Harish}\ and\ \citenamefont
  {Hansel}(2015)}]{harish2015asynchronous}%
  \BibitemOpen
  \bibfield  {author} {\bibinfo {author} {\bibfnamefont {O.}~\bibnamefont
  {Harish}}\ and\ \bibinfo {author} {\bibfnamefont {D.}~\bibnamefont
  {Hansel}},\ }\bibfield  {title} {\bibinfo {title} {Asynchronous rate chaos in
  spiking neuronal circuits},\ }\href@noop {} {\bibfield  {journal} {\bibinfo
  {journal} {PLoS Comput Biol}\ }\textbf {\bibinfo {volume} {11}},\ \bibinfo
  {pages} {e1004266} (\bibinfo {year} {2015})}\BibitemShut {NoStop}%
\bibitem [{\citenamefont {Stern}\ \emph {et~al.}(2014)\citenamefont {Stern},
  \citenamefont {Sompolinsky},\ and\ \citenamefont
  {Abbott}}]{stern2014dynamics}%
  \BibitemOpen
  \bibfield  {author} {\bibinfo {author} {\bibfnamefont {M.}~\bibnamefont
  {Stern}}, \bibinfo {author} {\bibfnamefont {H.}~\bibnamefont {Sompolinsky}},\
  and\ \bibinfo {author} {\bibfnamefont {L.}~\bibnamefont {Abbott}},\
  }\bibfield  {title} {\bibinfo {title} {Dynamics of random neural networks
  with bistable units},\ }\href@noop {} {\bibfield  {journal} {\bibinfo
  {journal} {Physical Review E}\ }\textbf {\bibinfo {volume} {90}},\ \bibinfo
  {pages} {062710} (\bibinfo {year} {2014})}\BibitemShut {NoStop}%
\bibitem [{\citenamefont {Rajan}\ \emph {et~al.}(2010)\citenamefont {Rajan},
  \citenamefont {Abbott},\ and\ \citenamefont
  {Sompolinsky}}]{rajan2010stimulus}%
  \BibitemOpen
  \bibfield  {author} {\bibinfo {author} {\bibfnamefont {K.}~\bibnamefont
  {Rajan}}, \bibinfo {author} {\bibfnamefont {L.}~\bibnamefont {Abbott}},\ and\
  \bibinfo {author} {\bibfnamefont {H.}~\bibnamefont {Sompolinsky}},\
  }\bibfield  {title} {\bibinfo {title} {Stimulus-dependent suppression of
  chaos in recurrent neural networks},\ }\href@noop {} {\bibfield  {journal}
  {\bibinfo  {journal} {Physical Review E}\ }\textbf {\bibinfo {volume} {82}},\
  \bibinfo {pages} {011903} (\bibinfo {year} {2010})}\BibitemShut {NoStop}%
\bibitem [{\citenamefont {Schuecker}\ \emph {et~al.}(2018)\citenamefont
  {Schuecker}, \citenamefont {Goedeke},\ and\ \citenamefont
  {Helias}}]{schuecker2018optimal}%
  \BibitemOpen
  \bibfield  {author} {\bibinfo {author} {\bibfnamefont {J.}~\bibnamefont
  {Schuecker}}, \bibinfo {author} {\bibfnamefont {S.}~\bibnamefont {Goedeke}},\
  and\ \bibinfo {author} {\bibfnamefont {M.}~\bibnamefont {Helias}},\
  }\bibfield  {title} {\bibinfo {title} {Optimal sequence memory in driven
  random networks},\ }\href@noop {} {\bibfield  {journal} {\bibinfo  {journal}
  {Physical Review X}\ }\textbf {\bibinfo {volume} {8}},\ \bibinfo {pages}
  {041029} (\bibinfo {year} {2018})}\BibitemShut {NoStop}%
\bibitem [{\citenamefont {Mastrogiuseppe}\ and\ \citenamefont
  {Ostojic}(2018)}]{mastrogiuseppe2018linking}%
  \BibitemOpen
  \bibfield  {author} {\bibinfo {author} {\bibfnamefont {F.}~\bibnamefont
  {Mastrogiuseppe}}\ and\ \bibinfo {author} {\bibfnamefont {S.}~\bibnamefont
  {Ostojic}},\ }\bibfield  {title} {\bibinfo {title} {Linking connectivity,
  dynamics, and computations in low-rank recurrent neural networks},\
  }\href@noop {} {\bibfield  {journal} {\bibinfo  {journal} {Neuron}\ }\textbf
  {\bibinfo {volume} {99}},\ \bibinfo {pages} {609} (\bibinfo {year}
  {2018})}\BibitemShut {NoStop}%
\bibitem [{\citenamefont {Landau}\ and\ \citenamefont
  {Sompolinsky}(2018)}]{landau2018coherent}%
  \BibitemOpen
  \bibfield  {author} {\bibinfo {author} {\bibfnamefont {I.~D.}\ \bibnamefont
  {Landau}}\ and\ \bibinfo {author} {\bibfnamefont {H.}~\bibnamefont
  {Sompolinsky}},\ }\bibfield  {title} {\bibinfo {title} {Coherent chaos in a
  recurrent neural network with structured connectivity},\ }\href@noop {}
  {\bibfield  {journal} {\bibinfo  {journal} {bioRxiv}\ ,\ \bibinfo {pages}
  {350801}} (\bibinfo {year} {2018})}\BibitemShut {NoStop}%
\bibitem [{\citenamefont {Landau}\ and\ \citenamefont
  {Sompolinsky}(2021)}]{landau2021macroscopic}%
  \BibitemOpen
  \bibfield  {author} {\bibinfo {author} {\bibfnamefont {I.~D.}\ \bibnamefont
  {Landau}}\ and\ \bibinfo {author} {\bibfnamefont {H.}~\bibnamefont
  {Sompolinsky}},\ }\bibfield  {title} {\bibinfo {title} {Macroscopic
  fluctuations emerge in balanced networks with incomplete recurrent
  alignment},\ }\href@noop {} {\bibfield  {journal} {\bibinfo  {journal}
  {Physical Review Research}\ }\textbf {\bibinfo {volume} {3}},\ \bibinfo
  {pages} {023171} (\bibinfo {year} {2021})}\BibitemShut {NoStop}%
\bibitem [{\citenamefont {Toyoizumi}\ and\ \citenamefont
  {Abbott}(2011)}]{toyoizumi2011beyond}%
  \BibitemOpen
  \bibfield  {author} {\bibinfo {author} {\bibfnamefont {T.}~\bibnamefont
  {Toyoizumi}}\ and\ \bibinfo {author} {\bibfnamefont {L.}~\bibnamefont
  {Abbott}},\ }\bibfield  {title} {\bibinfo {title} {Beyond the edge of chaos:
  Amplification and temporal integration by recurrent networks in the chaotic
  regime},\ }\href@noop {} {\bibfield  {journal} {\bibinfo  {journal} {Physical
  Review E}\ }\textbf {\bibinfo {volume} {84}},\ \bibinfo {pages} {051908}
  (\bibinfo {year} {2011})}\BibitemShut {NoStop}%
\bibitem [{\citenamefont {Sussillo}\ and\ \citenamefont
  {Abbott}(2009)}]{sussillo2009generating}%
  \BibitemOpen
  \bibfield  {author} {\bibinfo {author} {\bibfnamefont {D.}~\bibnamefont
  {Sussillo}}\ and\ \bibinfo {author} {\bibfnamefont {L.~F.}\ \bibnamefont
  {Abbott}},\ }\bibfield  {title} {\bibinfo {title} {Generating coherent
  patterns of activity from chaotic neural networks},\ }\href@noop {}
  {\bibfield  {journal} {\bibinfo  {journal} {Neuron}\ }\textbf {\bibinfo
  {volume} {63}},\ \bibinfo {pages} {544} (\bibinfo {year} {2009})}\BibitemShut
  {NoStop}%
\bibitem [{\citenamefont {Poole}\ \emph {et~al.}(2016)\citenamefont {Poole},
  \citenamefont {Lahiri}, \citenamefont {Raghu}, \citenamefont
  {Sohl-Dickstein},\ and\ \citenamefont {Ganguli}}]{poole2016exponential}%
  \BibitemOpen
  \bibfield  {author} {\bibinfo {author} {\bibfnamefont {B.}~\bibnamefont
  {Poole}}, \bibinfo {author} {\bibfnamefont {S.}~\bibnamefont {Lahiri}},
  \bibinfo {author} {\bibfnamefont {M.}~\bibnamefont {Raghu}}, \bibinfo
  {author} {\bibfnamefont {J.}~\bibnamefont {Sohl-Dickstein}},\ and\ \bibinfo
  {author} {\bibfnamefont {S.}~\bibnamefont {Ganguli}},\ }\bibfield  {title}
  {\bibinfo {title} {Exponential expressivity in deep neural networks through
  transient chaos},\ }\href@noop {} {\bibfield  {journal} {\bibinfo  {journal}
  {Advances in neural information processing systems}\ }\textbf {\bibinfo
  {volume} {29}},\ \bibinfo {pages} {3360} (\bibinfo {year}
  {2016})}\BibitemShut {NoStop}%
\bibitem [{\citenamefont {Keup}\ \emph {et~al.}(2021)\citenamefont {Keup},
  \citenamefont {K{\"u}hn}, \citenamefont {Dahmen},\ and\ \citenamefont
  {Helias}}]{keup2021transient}%
  \BibitemOpen
  \bibfield  {author} {\bibinfo {author} {\bibfnamefont {C.}~\bibnamefont
  {Keup}}, \bibinfo {author} {\bibfnamefont {T.}~\bibnamefont {K{\"u}hn}},
  \bibinfo {author} {\bibfnamefont {D.}~\bibnamefont {Dahmen}},\ and\ \bibinfo
  {author} {\bibfnamefont {M.}~\bibnamefont {Helias}},\ }\bibfield  {title}
  {\bibinfo {title} {Transient chaotic dimensionality expansion by recurrent
  networks},\ }\href@noop {} {\bibfield  {journal} {\bibinfo  {journal}
  {Physical Review X}\ }\textbf {\bibinfo {volume} {11}},\ \bibinfo {pages}
  {021064} (\bibinfo {year} {2021})}\BibitemShut {NoStop}%
\bibitem [{\citenamefont {Amit}\ and\ \citenamefont
  {Fusi}(1992)}]{amit1992constraints}%
  \BibitemOpen
  \bibfield  {author} {\bibinfo {author} {\bibfnamefont {D.~J.}\ \bibnamefont
  {Amit}}\ and\ \bibinfo {author} {\bibfnamefont {S.}~\bibnamefont {Fusi}},\
  }\bibfield  {title} {\bibinfo {title} {Constraints on learning in dynamic
  synapses},\ }\href@noop {} {\bibfield  {journal} {\bibinfo  {journal}
  {Network: Computation in Neural Systems}\ }\textbf {\bibinfo {volume} {3}},\
  \bibinfo {pages} {443} (\bibinfo {year} {1992})}\BibitemShut {NoStop}%
\bibitem [{\citenamefont {Amit}\ and\ \citenamefont
  {Huang}(2010)}]{amit2010precise}%
  \BibitemOpen
  \bibfield  {author} {\bibinfo {author} {\bibfnamefont {Y.}~\bibnamefont
  {Amit}}\ and\ \bibinfo {author} {\bibfnamefont {Y.}~\bibnamefont {Huang}},\
  }\bibfield  {title} {\bibinfo {title} {Precise capacity analysis in binary
  networks with multiple coding level inputs},\ }\href@noop {} {\bibfield
  {journal} {\bibinfo  {journal} {Neural computation}\ }\textbf {\bibinfo
  {volume} {22}},\ \bibinfo {pages} {660} (\bibinfo {year} {2010})}\BibitemShut
  {NoStop}%
\bibitem [{\citenamefont {Schuessler}\ \emph {et~al.}(2020)\citenamefont
  {Schuessler}, \citenamefont {Dubreuil}, \citenamefont {Mastrogiuseppe},
  \citenamefont {Ostojic},\ and\ \citenamefont
  {Barak}}]{schuessler2020dynamics}%
  \BibitemOpen
  \bibfield  {author} {\bibinfo {author} {\bibfnamefont {F.}~\bibnamefont
  {Schuessler}}, \bibinfo {author} {\bibfnamefont {A.}~\bibnamefont
  {Dubreuil}}, \bibinfo {author} {\bibfnamefont {F.}~\bibnamefont
  {Mastrogiuseppe}}, \bibinfo {author} {\bibfnamefont {S.}~\bibnamefont
  {Ostojic}},\ and\ \bibinfo {author} {\bibfnamefont {O.}~\bibnamefont
  {Barak}},\ }\bibfield  {title} {\bibinfo {title} {Dynamics of random
  recurrent networks with correlated low-rank structure},\ }\href@noop {}
  {\bibfield  {journal} {\bibinfo  {journal} {Physical Review Research}\
  }\textbf {\bibinfo {volume} {2}},\ \bibinfo {pages} {013111} (\bibinfo {year}
  {2020})}\BibitemShut {NoStop}%
\bibitem [{\citenamefont {Beiran}\ \emph {et~al.}(2021)\citenamefont {Beiran},
  \citenamefont {Dubreuil}, \citenamefont {Valente}, \citenamefont
  {Mastrogiuseppe},\ and\ \citenamefont {Ostojic}}]{beiran2021shaping}%
  \BibitemOpen
  \bibfield  {author} {\bibinfo {author} {\bibfnamefont {M.}~\bibnamefont
  {Beiran}}, \bibinfo {author} {\bibfnamefont {A.}~\bibnamefont {Dubreuil}},
  \bibinfo {author} {\bibfnamefont {A.}~\bibnamefont {Valente}}, \bibinfo
  {author} {\bibfnamefont {F.}~\bibnamefont {Mastrogiuseppe}},\ and\ \bibinfo
  {author} {\bibfnamefont {S.}~\bibnamefont {Ostojic}},\ }\bibfield  {title}
  {\bibinfo {title} {Shaping dynamics with multiple populations in low-rank
  recurrent networks},\ }\href@noop {} {\bibfield  {journal} {\bibinfo
  {journal} {Neural Computation}\ }\textbf {\bibinfo {volume} {33}},\ \bibinfo
  {pages} {1572} (\bibinfo {year} {2021})}\BibitemShut {NoStop}%
\bibitem [{\citenamefont {Tsodyks}(1988)}]{tsodyks1988associative}%
  \BibitemOpen
  \bibfield  {author} {\bibinfo {author} {\bibfnamefont {M.}~\bibnamefont
  {Tsodyks}},\ }\bibfield  {title} {\bibinfo {title} {Associative memory in
  asymmetric diluted network with low level of activity},\ }\href@noop {}
  {\bibfield  {journal} {\bibinfo  {journal} {EPL (Europhysics Letters)}\
  }\textbf {\bibinfo {volume} {7}},\ \bibinfo {pages} {203} (\bibinfo {year}
  {1988})}\BibitemShut {NoStop}%
\bibitem [{\citenamefont {Derrida}\ \emph {et~al.}(1987)\citenamefont
  {Derrida}, \citenamefont {Gardner},\ and\ \citenamefont
  {Zippelius}}]{derrida1987exactly}%
  \BibitemOpen
  \bibfield  {author} {\bibinfo {author} {\bibfnamefont {B.}~\bibnamefont
  {Derrida}}, \bibinfo {author} {\bibfnamefont {E.}~\bibnamefont {Gardner}},\
  and\ \bibinfo {author} {\bibfnamefont {A.}~\bibnamefont {Zippelius}},\
  }\bibfield  {title} {\bibinfo {title} {An exactly solvable asymmetric neural
  network model},\ }\href@noop {} {\bibfield  {journal} {\bibinfo  {journal}
  {EPL (Europhysics Letters)}\ }\textbf {\bibinfo {volume} {4}},\ \bibinfo
  {pages} {167} (\bibinfo {year} {1987})}\BibitemShut {NoStop}%
\bibitem [{\citenamefont {Huang}\ \emph {et~al.}(2018)\citenamefont {Huang},
  \citenamefont {Ramachandran}, \citenamefont {Lee},\ and\ \citenamefont
  {Olson}}]{huang2018neural}%
  \BibitemOpen
  \bibfield  {author} {\bibinfo {author} {\bibfnamefont {G.}~\bibnamefont
  {Huang}}, \bibinfo {author} {\bibfnamefont {S.}~\bibnamefont {Ramachandran}},
  \bibinfo {author} {\bibfnamefont {T.~S.}\ \bibnamefont {Lee}},\ and\ \bibinfo
  {author} {\bibfnamefont {C.~R.}\ \bibnamefont {Olson}},\ }\bibfield  {title}
  {\bibinfo {title} {Neural correlate of visual familiarity in macaque area
  v2},\ }\href@noop {} {\bibfield  {journal} {\bibinfo  {journal} {Journal of
  Neuroscience}\ }\textbf {\bibinfo {volume} {38}},\ \bibinfo {pages} {8967}
  (\bibinfo {year} {2018})}\BibitemShut {NoStop}%
\bibitem [{\citenamefont {Garrett}\ \emph {et~al.}(2020)\citenamefont
  {Garrett}, \citenamefont {Manavi}, \citenamefont {Roll}, \citenamefont
  {Ollerenshaw}, \citenamefont {Groblewski}, \citenamefont {Ponvert},
  \citenamefont {Kiggins}, \citenamefont {Casal}, \citenamefont {Mace},
  \citenamefont {Williford} \emph {et~al.}}]{garrett2020experience}%
  \BibitemOpen
  \bibfield  {author} {\bibinfo {author} {\bibfnamefont {M.}~\bibnamefont
  {Garrett}}, \bibinfo {author} {\bibfnamefont {S.}~\bibnamefont {Manavi}},
  \bibinfo {author} {\bibfnamefont {K.}~\bibnamefont {Roll}}, \bibinfo {author}
  {\bibfnamefont {D.~R.}\ \bibnamefont {Ollerenshaw}}, \bibinfo {author}
  {\bibfnamefont {P.~A.}\ \bibnamefont {Groblewski}}, \bibinfo {author}
  {\bibfnamefont {N.~D.}\ \bibnamefont {Ponvert}}, \bibinfo {author}
  {\bibfnamefont {J.~T.}\ \bibnamefont {Kiggins}}, \bibinfo {author}
  {\bibfnamefont {L.}~\bibnamefont {Casal}}, \bibinfo {author} {\bibfnamefont
  {K.}~\bibnamefont {Mace}}, \bibinfo {author} {\bibfnamefont {A.}~\bibnamefont
  {Williford}}, \emph {et~al.},\ }\bibfield  {title} {\bibinfo {title}
  {Experience shapes activity dynamics and stimulus coding of vip inhibitory
  cells},\ }\href@noop {} {\bibfield  {journal} {\bibinfo  {journal} {Elife}\
  }\textbf {\bibinfo {volume} {9}},\ \bibinfo {pages} {e50340} (\bibinfo {year}
  {2020})}\BibitemShut {NoStop}%
\bibitem [{\citenamefont {Amit}\ and\ \citenamefont
  {Brunel}(1997)}]{amit1997model}%
  \BibitemOpen
  \bibfield  {author} {\bibinfo {author} {\bibfnamefont {D.~J.}\ \bibnamefont
  {Amit}}\ and\ \bibinfo {author} {\bibfnamefont {N.}~\bibnamefont {Brunel}},\
  }\bibfield  {title} {\bibinfo {title} {Model of global spontaneous activity
  and local structured activity during delay periods in the cerebral cortex.},\
  }\href@noop {} {\bibfield  {journal} {\bibinfo  {journal} {Cerebral cortex}\
  }\textbf {\bibinfo {volume} {7}},\ \bibinfo {pages} {237} (\bibinfo {year}
  {1997})}\BibitemShut {NoStop}%
\bibitem [{\citenamefont {Wang}(1999)}]{wang1999synaptic}%
  \BibitemOpen
  \bibfield  {author} {\bibinfo {author} {\bibfnamefont {X.-J.}\ \bibnamefont
  {Wang}},\ }\bibfield  {title} {\bibinfo {title} {Synaptic basis of cortical
  persistent activity: the importance of nmda receptors to working memory},\
  }\href@noop {} {\bibfield  {journal} {\bibinfo  {journal} {Journal of
  Neuroscience}\ }\textbf {\bibinfo {volume} {19}},\ \bibinfo {pages} {9587}
  (\bibinfo {year} {1999})}\BibitemShut {NoStop}%
\bibitem [{\citenamefont {Lundqvist}\ \emph {et~al.}(2018)\citenamefont
  {Lundqvist}, \citenamefont {Herman},\ and\ \citenamefont
  {Miller}}]{lundqvist2018working}%
  \BibitemOpen
  \bibfield  {author} {\bibinfo {author} {\bibfnamefont {M.}~\bibnamefont
  {Lundqvist}}, \bibinfo {author} {\bibfnamefont {P.}~\bibnamefont {Herman}},\
  and\ \bibinfo {author} {\bibfnamefont {E.~K.}\ \bibnamefont {Miller}},\
  }\bibfield  {title} {\bibinfo {title} {Working memory: Delay activity, yes!
  persistent activity? maybe not},\ }\href@noop {} {\bibfield  {journal}
  {\bibinfo  {journal} {Journal of Neuroscience}\ }\textbf {\bibinfo {volume}
  {38}},\ \bibinfo {pages} {7013} (\bibinfo {year} {2018})}\BibitemShut
  {NoStop}%
\bibitem [{\citenamefont {Aljadeff}\ \emph {et~al.}(2021)\citenamefont
  {Aljadeff}, \citenamefont {Gillett}, \citenamefont {Pereira-Obilinovic},\
  and\ \citenamefont {Brunel}}]{aljadeff2021synapse}%
  \BibitemOpen
  \bibfield  {author} {\bibinfo {author} {\bibfnamefont {J.}~\bibnamefont
  {Aljadeff}}, \bibinfo {author} {\bibfnamefont {M.}~\bibnamefont {Gillett}},
  \bibinfo {author} {\bibfnamefont {U.}~\bibnamefont {Pereira-Obilinovic}},\
  and\ \bibinfo {author} {\bibfnamefont {N.}~\bibnamefont {Brunel}},\
  }\bibfield  {title} {\bibinfo {title} {From synapse to network: models of
  information storage and retrieval in neural circuits},\ }\href@noop {}
  {\bibfield  {journal} {\bibinfo  {journal} {Current opinion in neurobiology}\
  }\textbf {\bibinfo {volume} {70}},\ \bibinfo {pages} {24} (\bibinfo {year}
  {2021})}\BibitemShut {NoStop}%
\end{thebibliography}%

\end{document}